\documentclass[iop,twocolumn,tighten,numberedappendix,twocolappendix,revtex4]{emulateapj}
\usepackage[caption=false]{subfig}
\usepackage{amsmath}
\usepackage{amssymb}
\usepackage{lipsum}
\usepackage{xcolor}
\usepackage{soul} 

\graphicspath{{./}{plots/}}

\shorttitle{The Composition of the Ultra-faint Radio Population}
\shortauthors{Algera et al.}

\begin{document}

\title{A Multi-wavelength Analysis of the Faint Radio Sky (COSMOS-XS): the Nature of the Ultra-faint Radio Population}

\author{H. S. B. Algera\altaffilmark{1}}
\altaffiltext{1}{Leiden Observatory, Leiden University, P.O. Box 9513, 2300 RA Leiden, the Netherlands}
\email{algera@strw.leidenuniv.nl}

\author{D. van der Vlugt\altaffilmark{1}}

\author{J. A. Hodge\altaffilmark{1}} 

\author{I. Smail\altaffilmark{2}}
\altaffiltext{2}{Centre for Extragalactic Astronomy, Durham University, Department of Physics, South Road, Durham, DH1 3LE, UK}

\author{M. Novak\altaffilmark{3}}
\altaffiltext{3}{Max-Planck-Institut f\"ur Astronomie, K\"onigstuhl 17, D-69117 Heidelberg, Germany}

\author{J. F. Radcliffe\altaffilmark{4,5}}
\altaffiltext{4}{Department of Physics, University of Pretoria, Lynnwood Road, Hatfield, Pretoria 0083, South Africa}
\altaffiltext{5}{Jodrell Bank Centre for Astrophysics, The University of Manchester, SK11 9DL. United Kingdom}

\author{D. A. Riechers\altaffilmark{6,3}}
\altaffiltext{6}{Department of Astronomy, Cornell University, Space Sciences Building, Ithaca, NY 14853, USA}

\author{H. R\"{o}ttgering\altaffilmark{1}}

\author{V. Smol\v{c}i\'{c}\altaffilmark{7}}
\altaffiltext{7}{Department of Physics, University of Zagreb, Bijeni\^{c}ka cesta 32, 10002 Zagreb, Croatia}

\author{F. Walter\altaffilmark{3}}

\def\baselinestretch{1.0}


\begin{abstract}

Ultra-deep radio surveys are an invaluable probe of dust-obscured star formation, but require a clear understanding of the relative contribution from radio AGN to be used to their fullest potential. We study the composition of the $\mu$Jy radio population detected in the \emph{Karl G. Jansky} Very Large Array COSMOS-XS survey based on a sample of 1540 sources detected at 3 GHz over an area of $\sim350\,\text{arcmin}^2$. This ultra-deep survey consists of a single pointing in the well-studied COSMOS field at both 3 and 10 GHz and reaches RMS-sensitivities of $0.53$ and $0.41\,\mu$Jy beam$^{-1}$, respectively. We find multi-wavelength counterparts for $97\%$ of radio sources, based on a combination of near-UV/optical to sub-mm data, and through a stacking analysis at optical/near-infrared wavelengths we further show that the sources lacking such counterparts are likely to be high-redshift in nature (typical $z\sim4-5$). Utilizing the multi-wavelength data over COSMOS, we identify AGN through a variety of diagnostics and find these to make up $23.2\pm1.3\%$ of our sample, with the remainder constituting uncontaminated star-forming galaxies. However, more than half of the AGN exhibit radio emission consistent with originating from star-formation, with only $8.8\pm0.8\%$ of radio sources showing a clear excess in radio luminosity. At flux densities of $\sim30\,\mu$Jy at 3 GHz, the fraction of star-formation powered sources reaches $\sim90\%$, and this fraction is consistent with unity at even lower flux densities. Overall, our findings imply that ultra-deep radio surveys such as COSMOS-XS constitute a highly effective means of obtaining clean samples of star-formation powered radio sources.

\end{abstract}


\section{INTRODUCTION}
\label{sec:intro}
One of the key quests in extragalactic astronomy is to understand how the build-up and subsequent evolution of galaxies proceeds across cosmic time. Deep radio surveys offer an invaluable window onto this evolution, as they are a probe of both recent dust-unbiased star formation activity, as well as emission from active galactic nuclei (AGN). Past radio surveys were mostly limited to probing the latter, as AGN make up the bulk of the bright radio population (e.g.\ \citealt{condon1989}). Current surveys, in large part owing to the increased sensitivity of the upgraded NSF's \emph{Karl G. Jansky} Very Large Array (VLA), are now changing this, and allow for joint studies of star-forming galaxies (SFGs) and faint AGN. This, however, requires these populations be distinguished from each other, necessitating detailed studies of the multi-wavelength properties of the radio-detected population (e.g.\ \citealt{bonzini2013,padovani2017,smolcic2017b,delvecchio2017}).

Radio emission from SFGs is dominated at frequencies below $\lesssim30$ GHz by non-thermal synchrotron radiation \citep{condon1992}, which is thought to originate mainly from the shocks produced by the supernova explosions that end the lives of massive, short-lived stars. This conclusion is supported by the existence of the far-infrared-radio correlation (FIRC, e.g. \citealt{dejong1985,helou1985,yun2001,bell2003}), which constitutes a tight correlation between the (predominantly non-thermal) radio and far-infrared (FIR) emission of star-forming galaxies. As the FIR-emission is dominated by thermal radiation from dust that has been heated by young, massive stars, this allows for the usage of radio continuum emission as a star-formation tracer through the FIR-radio correlation. Radio observations of the star-forming population are therefore, by definition, dust-unbiased, and hence provide an invaluable probe of cosmic star formation. In particular, radio surveys directly complement rest-frame ultra-violet (UV) studies of the cosmic star formation rate density (SFRD), which, while extending out to very high redshift ($z\sim10$, e.g. \citealt{bouwens2015,oesch2018}), are highly sensitive to attenuation by dust. The extent of such dust attenuation remains highly uncertain beyond `cosmic noon' ($z\gtrsim3$, e.g. \citealt{casey2018}), and necessitates the additional use of dust-unbiased tracers of star formation. Recently, \citet{novak2017} performed the first study of the radio SFRD out to $z\sim5$, finding evidence that UV-based studies may underestimate the SFRD by $15-20\%$ beyond $z\gtrsim4$. They attribute this to substantial star formation occurring in dust-obscured galaxies, which further highlights the value in carrying out deep radio surveys. Their findings are consistent with results from large sub-millimeter surveys, which are predominantly sensitive to this dusty star-forming population (see \citealt{casey2014} for a review).

In the last few decades, it has also become increasingly clear that the evolution of individual galaxies is substantially affected by the presence of supermassive black holes (SMBHs) in their center (e.g.\ \citealt{kormendy2013}). Among such evidence found locally is the Magorrian-relation \citep{magorrian1998} describing the tight correlation between the mass of the central SMBH of a galaxy and of its bulge. In addition, the cosmic history of black hole growth is comparable to the growth in stellar mass \citep{shankar2009}, suggesting strong co-evolution. This is often explained by the accretion processes onto the black hole regulating star formation through mechanical feedback, either impeding (e.g.\ \citealt{best2006,farrah2012}) or instead triggering epochs of star formation and (\citealt{wang2010,reines2011}). Such AGN feedback is vital in particular for numerical simulations (e.g.\ \citealt{springel2005,schaye2015}) in order to recover, for example, galaxy luminosity functions and local scaling relations. Furthermore, direct evidence for mechanical feedback has been observed in local systems (e.g.\ \citealt{mcnamara2012,morganti2013}). These results exemplify the importance of studying both galaxies and AGN jointly, instead of as separate entities. 

Typically, two populations of AGN can be distinguished in radio surveys: sources which can be identified through an excess in radio emission compared to what is expected from the FIRC (henceforth referred to as radio-excess AGN) and sources which have radio emission compatible with originating from star formation, but can be identified as AGN through any of several multi-wavelength diagnostics \citep{padovani2011,bonzini2013,heckmanbest2014,delvecchio2017,smolcic2017b,calistrorivera2017}.\footnote{These sources are also often referred to as radio-loud and radio-quiet AGN respectively (e.g.\ \citealt{heckmanbest2014}), though we follow here the terminology from \citet{delvecchio2017} and \citet{smolcic2017b}.} The latter class generally exhibit AGN-related emission throughout the bulk of their non-radio spectral energy distribution (SED), in the form of, e.g., strong X-ray emission or mid-infrared dust emission from a warm torus surrounding the AGN (e.g. \citealt{evans2006,hardcastle2013}). This distinction illustrates the importance of using multi-wavelength diagnostics for identifying AGN activity, which forms the focus of this work.

The distribution of the star-forming and AGN populations has been well-established to be a strong function of radio flux density. At high flux densities, the radio-detected population is dominated by radio-excess AGN \citep{kellermann1987,condon1989}, followed by a flattening of the number counts around $\sim1$ mJy. This flattening was initially interpreted as the advent of purely star-forming galaxies, but subsequent studies (e.g. \citealt{gruppioni1999,bonzini2013,padovani2015}) probing down to a few hundreds to tens of $\mu$Jy at 1.4 GHz revealed that a substantial part of the sub-mJy population remains dominated by non-radio excess AGN, with the current consensus being that SFGs only start dominating the radio source counts below $\sim100\ \mu$Jy \citep{padovani2011,bonzini2013,padovani2015,maini2016,smolcic2017b}, which is also in fairly good agreement with predictions from semi-empirical models of the radio source counts \citep{wilman2008,bonaldi2018}. This faint regime is of great interest for studies of star-formation, but has not been widely accessible yet to present-day radio telescopes. 

With the upgraded VLA in particular, the radio population at the $\mu$Jy level can now reliably be probed, which will help constrain the relative contributions of the various radio populations to unprecedented flux densities. Historically, a `wedding-cake approach' has been the tried-and-tested design for radio surveys, incorporating both large-area observations and deeper exposures of smaller regions of the sky. By combining such observations, a clear consensus on both the bright and faint end of the radio population can be reached, which is crucial for understanding the different classes of radio-detected galaxies, as well as for the accurate determination of the radio source counts, luminosity functions and the subsequent radio-derived cosmic star-formation history.

The COSMOS-XS survey (Van der Vlugt et al. 2020, henceforth Paper I) was designed to explore the faint radio regime and, by construction, constitutes the top of the wedding cake, making it the natural complement to large-area surveys such as the 3 GHz VLA-COSMOS project \citep{smolcic2017b,smolcic2017a}. By going a factor of $\sim5$ deeper than this survey, we directly probe two of the most interesting populations, namely the poorly-understood radio-quiet AGN and the faint star-forming galaxies. While radio surveys were historically limited by their inability to probe the typical star-forming population, being sensitive mostly to starburst galaxies (typical star-formation rates in excess of $\gtrsim 100\,M_\odot\,\text{yr}^{-1}$), the COSMOS-XS survey reaches sub-$\mu$Jy depths and allows for the detection of typical star-forming sources out to redshifts of $z\lesssim3$. For this reason, the survey is well-suited to bridge the gap between the current deepest radio surveys and those of the next-generation radio telescopes.

COSMOS-XS targets a region of the well-studied COSMOS field \citep{scoville2007}, such that a wealth of multi-wavelength data are available for optimal classification of the radio population. Such ancillary data is crucial to place the survey into a wider astronomical context, and allows for the connection of the radio properties of the observed galaxy population to observations in the rest of the electromagnetic spectrum. With the combined COSMOS-XS survey and multi-wavelength data, we ultimately aim to constrain the faint end of the high-redshift radio luminosity functions of both SFGs and AGN, and use these to derived the corresponding dust-unbiased cosmic star formation history, as well as the AGN accretion history. Additionally, the unprecedented depth at multiple radio frequencies allows for the study of the high-redshift radio spectral energy distribution in great detail, and allows for the systematic isolation of the radio free-free component in star-forming galaxies, to be used as a robust star formation rate tracer (e.g \citealt{murphy2012,murphy2017}). All of these science goals require a good understanding of the origin of the observed radio emission and hence depend on whether it emanates from star formation or instead from an AGN, and will be tackled in forthcoming papers.

In this paper, the second in the series describing the COSMOS-XS radio survey, we address this question by studying the composition of the ultra-faint radio population detected at 3 GHz in an ultra-deep single pointing over the COSMOS field (reaching an RMS of $\sim0.53\,\mu$Jy beam$^{-1}$). We additionally present a catalog containing the multi-wavelength counterparts of the radio sources selected at 3 GHz, utilizing the wealth of X-ray to radio data available over COSMOS. The paper is structured as follows. In Section \ref{sec:data} we summarize the COSMOS-XS observations, the radio-selected catalog as well as the multi-wavelength ancillary data. In Section \ref{sec:matching} we describe the association of counterparts to the radio sample. The decomposition of the radio population into SFGs and AGN is laid out in Section \ref{sec:agn}, and we present the multi-wavelength counterpart catalog in Section \ref{app:catalog}. Finally, we present the ultra-faint radio soure counts, separated into SFGs and AGN, in Section \ref{sec:composition} and we summarize our findings in Section \ref{sec:summary}. Throughout this work, we assume a flat $\Lambda$-Cold Dark Matter cosmology with $\Omega_m=0.30$, $\Omega_\Lambda=0.70$ and $H_0=70$\,km\,s$^{-1}$\,Mpc$^{-1}$. Magnitudes are expressed in the AB system \citep{oke1983}, and a \citet{chabrier2003} initial mass function is assumed. The radio spectral index, $\alpha$, is defined through $S_\nu \propto \nu^\alpha$ where $S_\nu$ represents the flux density at frequency $\nu$.

\section{DATA}
\label{sec:data}

\subsection{Radio Data}
\label{sec:radiodata}
The COSMOS-XS survey consists of a single ultra-deep VLA pointing in the well-studied COSMOS field at both 3 and 10 GHz of $\sim100$ and $\sim90$ h of observation time, respectively. The full survey is described in detail in Paper I, but we summarize the key procedures and parameters here. The 3 GHz observations (also known as the S-band) were taken in B-array configuration, and span a total bandwidth of 2 GHz. The effective area of the pointing -- measured up to 20 per cent of the peak primary beam sensitivity at the central frequency -- is approximately $350$ arcmin$^2$. The 10 GHz pointing (X-band) was taken in C-array configuration, and spans a frequency range of 4 GHz around the central frequency. The total survey area is approximately $30\,$arcmin$^2$ at 10 GHz. At both frequencies, roughly 20\% of the bandwidth was lost due to excessive radio frequency interference.

Imaging of both datasets was performed using standalone imager {\sc{WSclean}} \citep{offringa2012}, incorporating $w-$stacking to account for the non-coplanarity of our baselines. Both images were created via Briggs weighting, with a robust parameter set to $0.5$. This resulted in a synthesized beam of $2\farcs14\times1\farcs81$ at 3 GHz and $2\farcs33\times2\farcs01$ at 10 GHz. The near-equal resolution of $\sim2\farcs0$ was chosen to be large enough to avoid resolving typical faint radio sources, which are generally sub-arcsecond in size at the $\mu$Jy-level \citep{bondi2018,cotton2018}. In turn, this resolution allows for the cleanest measurement of their radio flux densities, while ensuring that confusion noise remains negligible (Paper I).

The final root-mean-square (RMS) noise levels of the S- and X-band images are $0.53\ \mu\text{Jy beam}^{-1}$ and $0.41\ \mu\text{Jy beam}^{-1}$ at their respective pointing centres. For the S-band, the RMS-noise corresponds to a brightness temperature of $T_B \simeq 20\,$mK. This implies we are sensitive even to face-on star-forming spiral galaxies, which have a typical $T_B = 0.75\pm 0.25\,$K \citep{hummel1981}.

\begin{figure}[]
	\centering
	\hspace*{-0.3cm}\includegraphics[width=0.5\textwidth]{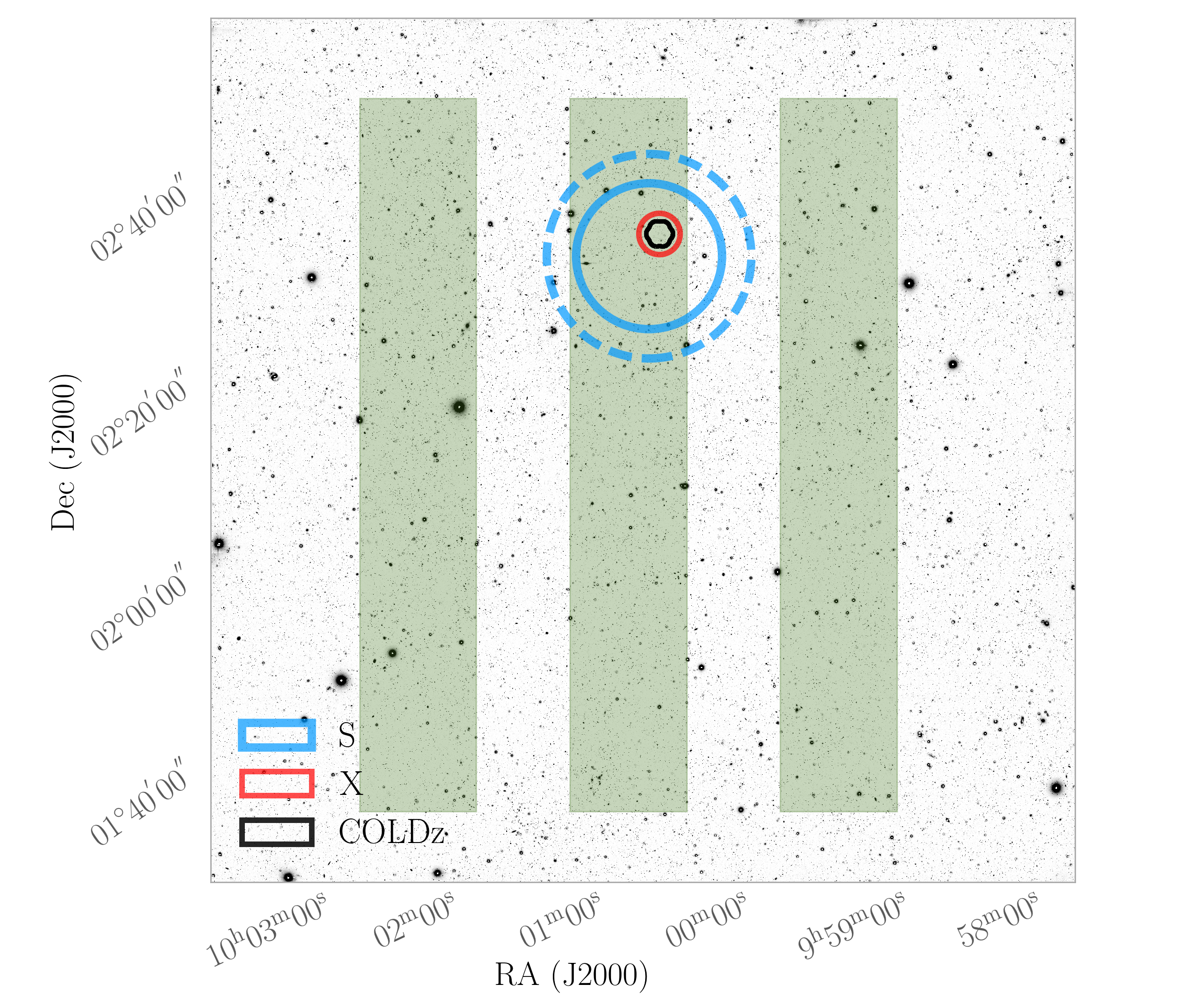}
	\caption{Layout of the S- (blue) and X-band (red) pointings, out to the half-power point of the primary beam (solid circles). For the S-band, we further highlight the field-of-view out to 20 per cent of the peak primary beam sensitivity (dashed circle), which defines our total survey area, over the two square degree COSMOS field. Also shown is the 34 GHz COLD$z$ mosaic \citep{pavesi2018}, which overlaps in its entirety with the X-band field of view. The background image consists of Subaru observations in the $i+$ filter, with the UltraVISTA ultra-deep stripes overlaid in green. For a zoomed-in view of the radio maps themselves, see Van der Vlugt et al.\ (2020).}
	\label{fig:cosmos}
\end{figure}

The locations of both the S- and X-band pointings within the COSMOS field are shown in Figure \ref{fig:cosmos}, and were explicitly chosen to match the pointing area of the COLD$z$ survey \citep{pavesi2018,riechers2019}. A zoomed-in view of the radio maps themselves are presented in Paper I. 

Source extraction on both images was performed with {\sc{PyBDSF}} \citep{mohanrafferty2015}, down to a $5\sigma$ peak flux threshold. {\sc{PyBDSF}} operates through identifying islands of contiguous emission around this peak value, and fitting such islands with elliptical Gaussians to obtain peak and integrated flux measurements. In total, we obtain 1498 distinct radio sources within 20\% of the peak primary beam sensitivity, of which 70 ($\sim5\%$) consist of multiple Gaussian components. While our survey is far from being confusion limited, a fraction of islands deemed robustly resolved in the source detection are in fact artificially extended as a result of source blending. In order to disentangle true extended sources from blended ones, we examined whether the Gaussian components making up an island can be individually cross-matched to separate sources in the recent Super-deblended catalog over COSMOS (\citealt{jin2018}, see Section \ref{sec:optidata}), which contains mid-IR to radio photometry based on positional priors from a combination of $K_s$, \emph{Spitzer}/MIPS $24\mu$m and VLA 1.4 and 3 GHz observations. As these radio data are both shallower and higher resolution than our observations, they are suitable for assessing any source blending. Furthermore, radio sources such as FR-II AGN \citep{fanaroffriley1974} are not expected to have mid-IR counterparts for their individual lobes as these sources $-$ by definition $-$ have their radio emission spatially offset from the host galaxy. Hence, when all components can be individually associated to different multi-wavelength counterparts, we deem these associations robust, and define the Gaussian components to be separate radio sources. Altogether, we find that 40 of the initial 70 multi-Gaussian sources separate into 82 single `deblended' components. In total, our radio survey therefore consists of 1540 individual sources detected at 3 GHz, within 20 per cent of the peak primary beam sensitivity. Altogether, this results in a $\sim4\%$ increase in the number of radio sources that is cross-matched to a multi-wavelength counterpart (Section \ref{sec:matching}). This is the result of two effects: firstly, we find an overall larger number of radio sources now that blended ones have been separated, and secondly we find a larger fractional number of cross-matches, as blended sources were assigned a flux-weighted source center that could be substantially offset from the true centers of the individual Gaussian components, preventing reliable cross-matching. We note that this procedure slightly deviates from that in Paper I, where we focused instead on the radio properties of this sample and therefore refrained from invoking multi-wavelength cross-matching.

In addition to the 1540 S-band detections, a total of 90 sources are detected at 10 GHz within $20\%$ of the maximum primary beam sensitivity at $\geq5\sigma$ significance (948 and 60 sources lie within the half power point of the primary beam, respectively).\footnote{With the adopted primary beam cut-off of 20\%, the COSMOS-XS survey is deeper than the ($\sim5\times$ shallower) 3 GHz VLA-COSMOS survey \citep{smolcic2017b,smolcic2017a} across the entire field of view.} The S-band sample comprises the main radio sample used in the subsequent analysis described in this paper. We assign the radio sources detected at 3 GHz either their peak or integrated flux density, following the method from \citet{bondi2008} described in detail in Paper I. We then ensure that we take the same flux density measurement for the X-band sources, as all sources detected at 10 GHz can be cross-matched to an S-band counterpart (Section \ref{sec:matching}) and the radio data have a similar resolution at both frequencies. In Paper I, we additionally investigated the completeness and reliability of the radio sample, for which we repeat the main conclusions here. We found that the catalogues are highly reliable with a low number of possible spurious detections ($\lesssim2\%$). We further investigate possible spurious sources by visually inspecting all detections within $30''$ of a bright ($\text{SNR} \geq 200$) radio source. From these we flag, but do not remove, eight sources that are potentially spurious or have their fluxes affected as result of the characteristic VLA dirty beam pattern around the nearby bright object. The S-band sample was further determined to be $\gtrsim90\%$ complete above integrated flux densities of $15\,\mu$Jy, with the completeness dropping to $50\%$ at $\sim10\,\mu$Jy due mainly to the primary beam attenuation reducing the survey area. In our derivation of the radio number counts for star-forming sources and AGN (Section \ref{sec:composition}) all of these completeness considerations are taken into account.

Additional radio data over our pointings exists as part of the 1.4 GHz VLA-COSMOS survey \citep{schinnerer2007,schinnerer2010}, which reaches an RMS of $\sim12\,\mu$Jy beam$^{-1}$. Accounting for the frequency difference through scaling with $\alpha=-0.70$, our S-band observations are a factor $\sim13$ deeper than these lower frequency data, and hence the 1.4 GHz observations are mostly useful for the brightest sources detected at 3 and 10 GHz. Additionally, a seven-pointing mosaic at 34 GHz ($\text{RMS}\sim1.4\mu$Jy beam$^{-1}$, area $\sim10\text{arcmin}^2$) exists as part of the COLDz project (\citealt{pavesi2018,riechers2019}; Algera et al.\ in prep.). The COSMOS-XS 10 GHz data is directly centered on this mosaic, allowing for a detailed analysis of the long-wavelength spectrum of faint radio sources with up to four frequencies. We defer this analysis to a future paper.

\subsection{Near-UV to far-IR data}\label{sec:optidata}
The COSMOS field has been the target of a considerable number of studies spanning the full electromagnetic spectrum. We complement our radio observations with near-UV to FIR-data that has been compiled into various multi-wavelength catalogs: i) the Super-deblended mid- to far-infrared catalog \citep{jin2018} containing photometry ranging from IRAC 3.6$\mu$m to 20cm (1.4 GHz) radio observations; ii) the $z^{++}YJHK_s$-selected catalog compiled by \citet{laigle2016} (hereafter COSMOS2015) and iii) the $i-$band selected catalog by \citet{capak2007}. 

The Super-deblended catalog contains the latest MIR-radio photometry for nearly 200,000 sources in the COSMOS field, with 12,335 located within the COSMOS-XS field-of-view. Due to the relatively poor resolution of FIR telescopes such as \emph{Herschel}, blending of sources introduces complications for accurate photometry. Through the use of priors on source positions from higher resolution images (\emph{Spitzer}/MIPS 24$\mu$m and VLA 1.4 and 3 GHz observations) in combination with point spread function fitting, the contributions to the flux from various blended galaxies in a low-resolution image can be partly disentangled. This `Super-deblending' procedure is described in detail in \citet{liu2018}. The Super-deblended catalog contains photometry from \emph{Spitzer}/IRAC \citep{sanders2007} and \emph{Spitzer}/MIPS $24\mu$m \citep{lefloch2009} as part of the S-COSMOS survey, \emph{Herschel}/PACS $100\mu$m and $160\mu$m data from the PEP \citep{lutz2011} and CANDELS-\emph{Herschel} (PI: M. Dickinson) programs, \emph{Herschel}/SPIRE images at $250, 350$ and $500\mu$m as part of the HerMES survey \citep{oliver2012} and further FIR data at 850$\mu$m from SCUBA2 as part of the Cosmology Legacy Survey \citep{geach2017}, AzTEC 1.1mm observations \citep{aretxaga2011} and MAMBO 1.2mm images \citep{bertoldi2007}, in addition to 1.4 GHz and 3 GHz radio observations from \citet{schinnerer2007,schinnerer2010} and \citet{smolcic2017a}, respectively. However, we use the photometry from catalogs provided directly by \citet{schinnerer2007,schinnerer2010}, as they provide both peak and integrated fluxes, whereas the Super-deblended catalog solely provides peak values. In addition, we use our $\sim5$ times deeper COSMOS-XS 3 GHz observations described in Section \ref{sec:radiodata} in favor of those from \citet{smolcic2017a}.

The Super-deblended catalog further includes photometric and spectroscopic redshifts, based on the \citet{laigle2016} catalog where available, in addition to IR-derived photometric redshifts and star-formation rates.\footnote{We do not use FIR-derived photometric redshifts in this work, as these values are considerably more uncertain than those derived from near-UV to near-IR photometry (e.g. \citealt{simpson2014}), but we comment on the small sample of sources with only FIR photometric redshifts in Section \ref{sec:radioonly}.}

Since a shallower radio catalog was used for the deblending procedure, this raises the concern that one of the VLA-COSMOS priors for the Super-deblended catalog is, in fact, located near a fainter radio source detected only in COSMOS-XS that contributes partially to the FIR flux at that location. In such a scenario, all the FIR emission would be wrongfully assigned to the brighter radio source, which would have an artificially boosted flux, and the fainter source may be wrongfully assigned no FIR-counterpart. We verified, however, that this is not likely to be an issue, as we see no drop in the fraction of cross-matches between COSMOS-XS and the Super-deblended catalog for radio sources with a nearby neighbor in COSMOS-XS. Hence there is no indication of any boosting in the FIR-fluxes of Super-deblended entries due to a nearby, faint radio source. \\

The COSMOS2015 catalog contains photometry for upwards of half a million entries over the 2 square-degree COSMOS field, including 37,841 within the COSMOS-XS field-of-view. Sources are drawn from a combined $z^{++}YJHK_s$ detection image, where the deep $YJHK_s$ observations are taken from the second UltraVISTA data release \citep{mccracken2012}. The COSMOS-XS S-band pointing is largely located within one of the UltraVISTA `ultra-deep' stripes (Figure \ref{fig:cosmos}), which reaches a $3\sigma$ depth in magnitude of $25.8, 25.4, 25.0, 25.2$ in $Y,J,H$ and $K_s$ respectively, as measured in $3''$ apertures. The COSMOS2015 catalog further provides cross-matches with NUV, and MIR/FIR data. The former consists of {\it{GALEX}} observations at 1500 \AA\ (FUV) and 2500 \AA\ (NUV) \citep{zamojski2007}, and the latter ranges from \emph{Spitzer}/MIPS $24\mu$m to \emph{Herschel}/SPIRE $500\mu$m photometry, drawn from the same programs as introduced for the Super-deblended catalog. In addition, photometric redshifts, star-formation rates and stellar masses are provided by COSMOS2015, derived through SED-fitting by {\sc{LePhare}} \citep{ilbert2006}, using photometry spanning the NUV - $K_s$ bands.

Finally, the $i-$band selected catalog compiles photometry from 15 photometric bands between the $u$-band ($0.3\mu$m) and the $2.5\mu$m $K_s$-band. The $5\sigma$ depth in the $i-$band equals 26.2 as determined within a $3''$ aperture. Photometric redshifts were derived from SED fitting with {\sc{LePhare}} and were later added to the catalog by \citet{ilbert2009}.

\subsection{Spectroscopic Redshifts}
A substantial fraction of galaxies in the COSMOS field have been targeted spectroscopically, and therefore have a robustly determined redshift. We make use of the `master spectroscopic redshift catalog' available internally to the COSMOS team (version 01 Sept. 2017, M. Salvato in prep.). It contains $\sim$100,000 entries, compiled from a large number of spectroscopic surveys over the COSMOS field, including zCOSMOS \citep{lilly2007,lilly2009}, the VIMOS Ultra Deep Survey (VUDS; \citealt{lefevre2015}) and MOSDEF \citep{kriek2015}. In total, there are 5,074 sources with reliable spectroscopic redshifts in the COSMOS-XS field-of-view (see also Section \ref{sec:specz}).

\subsection{X-ray data}
Strong X-ray emission is a vital diagnostic for AGN activity. The most recent X-ray data over the COSMOS field is the 4.6 Ms \emph{Chandra} COSMOS Legacy survey \citep{civano2016}, covering the full 2.2 deg$^2$. \citet{marchesi2016} present the catalog of the optical and infrared counterparts of the X-ray sources identified by the survey. This catalog includes 200 sources within our S-band field of view, and contains for each the absorption-corrected luminosity in the soft $[0.5 - 2]$ keV, hard $[2 - 10]$ keV and full $[0.5 - 10]$ keV bands, or the corresponding upper limit in the case of a non-detection in a given energy band.\footnote{The flux limit of the \emph{Chandra} COSMOS Legacy Survey over the COSMOS-XS S-band pointing is $\sim 2 \times 10^{-15}$ erg cm$^{-2}$ s$^{-1}$ in the full $2-10$ keV range \citep{civano2016}.} Where available, X-ray sources in this catalog were assigned the spectroscopic redshift of their counterparts, taken from the COSMOS master spectroscopic catalog. For the remaining sources, photometric redshifts exist, based on template fitting making use of AGN-specific templates as described in \citet{salvato2011}. X-ray luminosities were calculated by using the best available redshift and an X-ray spectral index of $\Gamma = 1.4$ for the required $K-$corrections, which is a typical value for a mix of obscured and onobscured sources \citep{marchesi2016b}. Absorption corrections to the X-ray luminosity in each energy band were calculated based on the measured hardness ratio, as described in \citet{civano2016}.


\section{MULTI-WAVELENGTH CROSS-MATCHING}
\label{sec:matching}

In this section, we elaborate on our multi-wavelength matching process. As a brief summary of the procedure, we cross-match catalogs based on a symmetric nearest neighbour algorithm, whereby we search for counterparts within a given matching radius. A suitable value of this radius is determined through cross-matching with a mock version of the appropriate catalog, which contains the same sources with randomized sky coordinates. As our radio images are not uniformly sensitive across their full field of view as result of the primary beam, we ensure mock sources can only be placed in the region where they can theoretically be detected at $\geq5\sigma$, to mimic the true distribution of sources. Through cross-matching with such mock catalogs, we obtain an estimate of the number of false matches at any given matching radius, and we hence define the false matching rate (FMR) as the average number of cross-matches obtained with the randomized catalogs divided by the total number of catalog entries. The matching radius we adopt is taken to be the radius where the number of matches for the real and mock catalogs is (approximately) equal, which is generally around $0\farcs9$, and coincides with a typical $\text{FMR}\lesssim3\%$ for all our multi-wavelength cross-matching.

\begin{figure}
    \centering
    \includegraphics[width=0.5\textwidth]{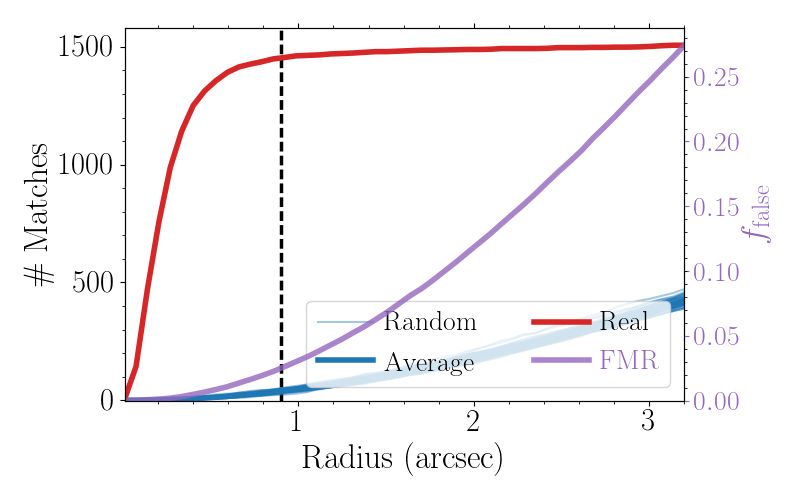}
    \caption{Illustration of the cross-matching process, for the radio and Super-deblended catalogs. The red line represents the number of matches obtained within a given matching radius, whereas the blue lines show the number of matches obtained when cross-matching with a catalog with the same source density but randomized sky positions. The purple curve, corresponding to the right ordinate axis, shows the false match fraction obtained at a given matching radius. The vertical black line indicates the typical matching radius of $0\farcs9$ adopted in this work.}
    \label{fig:matching}
\end{figure}

\subsection{Radio Cross-Matching}
There are a total of 1540 and 90 radio sources within $20\%$ of the peak primary beam sensitivity for the S- and X-band, respectively. We cross-match these two frequencies using a matching radius of $0\farcs9$, which yields 89 matches (FMR $\lesssim0.7\%$). The single X-band source that could not be cross-matched to a counterpart at 3 GHz appears to be a lobed radio source where the relative brightness of the two lobes is different in the two images, causing the centre of the source to be appreciably offset between the two images ($\sim1\farcs25$). Despite this offset, visual inspection verifies that the sources are related, such that all X-band sources are assigned a counterpart at 3 GHz. Due to the relatively low density of radio sources in the VLA COSMOS 1.4 GHz catalog, we utilize a matching radius of $1\farcs2$ when cross-matching with the S-band data ($\text{FMR}\lesssim 0.1\%$). This generates 185 matches, with 12 sources being detected at all three frequencies (1.4, 3 and 10 GHz). 

\subsection{Additional Cross-Matching}
In order to construct UV/optical -- FIR SEDs for the S-band detected sources, we cross-match with three catalogs in order of decreasing priority: Super-deblended \citep{jin2018}, COSMOS2015 \citep{laigle2016} and the $i-$band selected catalog by \citet{capak2007}. The Super-deblended catalog contains the most up-to-date collection of FIR-photometry available over the COSMOS field, but does not include optical and near-IR photometry shortwards of IRAC $3.6\mu$m. We therefore attempt to cross-match S-band sources that have Super-deblended counterparts with either COSMOS2015 or the $i-$band selected catalog, which do contain photometry at these shorter wavelengths. 

\begin{figure*}
	\centering
	\includegraphics[width=\textwidth]{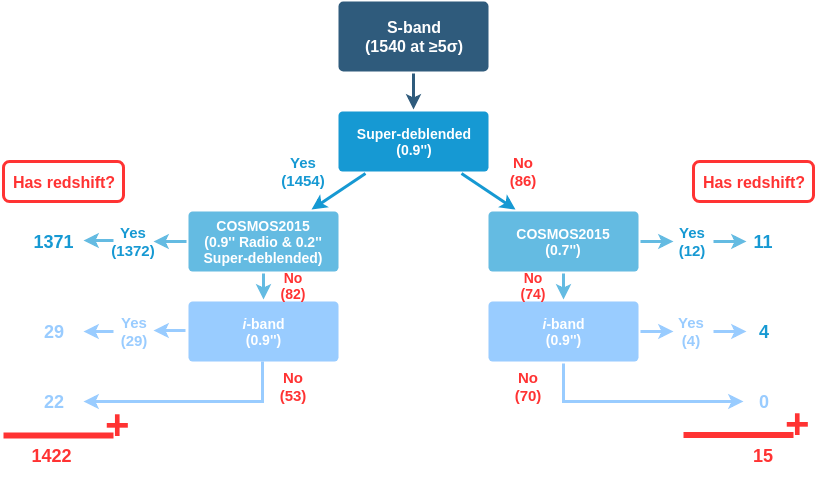}
	\caption{Flowchart of the matching process, indicating the priority of catalogs used and number of cross-matches obtained with each. We match a total of 1470/1540 sources (95.5\%) to a counterpart in at least one multi-wavelength catalog. From this sample, we keep 1437 sources (93.3\%) with reliable redshift information. This represents the main radio sample studied in this work.}
	\label{fig:flowchart}
\end{figure*}

An overview of the matching process is presented in the form of a flowchart in Figure \ref{fig:flowchart}. In summary, the procedure is as follows: we first cross-match the S-band selected catalog of 1540 sources with the Super-deblended photometric catalog, obtaining 1454 matches within $0\farcs9$ ($\text{FMR}\simeq2.5\%$, see Figure \ref{fig:matching}). To obtain optical and near-IR photometry, we subsequently cross-match with the COSMOS2015 catalog. As a result of the larger source density compared to the Super-deblended catalog, we use a matching radius of $0\farcs7$ instead of $0\farcs9$ around the radio coordinates, motivated by the number of false matches obtained at larger radii. The theoretical FMR at $0\farcs7$ equals $\sim4.8\%$, which is substantial. To account for both this and the difference in matching radii between the different catalogs, which may lead to inconsistencies in the overall cross-matching process where we artificially cannot assign our sources COSMOS2015 counterparts if their offset from the radio coordinates falls into the range $0\farcs7 < r < 0\farcs9$, we therefore also utilize the Super-deblended source coordinates. As these are based in part on IR-detections, they are typically more similar to the near-IR derived COSMOS2015 coordinates. We therefore perform additional cross-matching within $0\farcs2$ around these Super-deblended source positions $-$ since such closely associated sources are likely to be real ($\text{FMR} \lesssim 0.4\%$) $-$ under the constraint that the offset between any of the catalog is less than $0\farcs9$. We then acquire 1372 sources with both Super-deblended and COSMOS2015 photometry, which spans the full near-UV to millimetre range. Only 25 of these cross-matches ($1.8\%$) have COSMOS2015 - Super-deblended separations $\geq0\farcs2$, but as the median separation between the radio and Super-deblended coordinates remains small ($\sim0\farcs3$), we expect a negligible increase in the overall FMR after cross-matching with the COSMOS2015 catalog. We additionally have 82 sources with Super-deblended cross-matches for which we did not obtain COSMOS2015 counterparts. As these sources will greatly benefit from shorter wavelength data in our subsequent analysis, we cross-match them with the $i$-band selected catalog within $0\farcs9$ and recover an additional 29 matches.\footnote{The formal cross-matching radius adopted is $0\farcs9$ for consistency with other catalogs, but all matches lie within $0\farcs5$ of both the Super-deblended and radio sky positions, and as such we estimate the FMR to be $\lesssim1\%$.} We thus obtain full near-UV to radio photometry for $96.4\%$ of sources cross-matched with the Super-deblended catalog.

For the 86 ($5.6\%$) of radio-detected sources for which we could not acquire robust Super-deblended counterparts, we first cross-match with the COSMOS2015 catalog within $0\farcs7$, gaining 12 matches (4 expected false matches). For the remaining sources, we obtain 4 matches with the $i$-band catalog within $0\farcs9$, containing photometry up to the $K_s$-band (with 1 expected false match). Overall, we did not obtain any non-radio counterparts for 70/1540 sources ($4.5\%$). Two of these are detected at multiple radio frequencies and are therefore certainly real. Upon cross-matching with the S-COSMOS IRAC catalog from \citet{sanders2007}, we recover 18 additional matches within $0\farcs9$, which further indicates that a substantial number of this sample consists of real radio sources that evade detection at shorter wavelengths. However, as these cross-matches therefore solely have IRAC photometry, they lack redshift information, which is crucial for subsequent AGN identification (Section \ref{sec:agn}). Therefore, we do not further include these sources in the characterization of our radio sample, though we return to these `optically dark' detections in Section \ref{sec:radioonly}. When accounting for these additional 18 matches with S-COSMOS, $96.6\%$ of our radio sample can be cross-matched to a non-radio counterpart.

As reliable redshift information is crucial for investigating the physical properties of our radio-detected sources, we further remove 33 galaxies from our sample for which no redshift information was available in any of the catalogs. The majority of these sources are only present in the Super-deblended catalog, and were detected through priors from the 3 GHz VLA COSMOS project \citep{smolcic2017a}. As for these sources no photometry exists shortwards of MIPS $24\mu$m, no reliable photometric redshift can be obtained. Altogether, we are left with a remainder of 1437 sources ($93.3\%$ of the initial 1540 detections at 3 GHz). This constitutes the S-band detected sample that will be used in the subsequent analysis.

We additionally recover 108 cross-matches with X-ray sources taken from the \emph{Chandra} COSMOS Legacy survey \citep{civano2016,marchesi2016} based on an adopted cross-matching radius of $1\farcs4$ (theoretical $\text{FMR} \simeq 0.1\%$).\footnote{We adopted a cross-matching radius of $1\farcs4$ as \emph{Chandra} astrometry is accurate to 99\% within this radius, see \url{http://cxc.harvard.edu/cal/ASPECT/celmon/}.} The 108 X-ray counterparts for the radio sample correspond to $\sim51\%$ of the X-ray sources within the COSMOS-XS field of view. This is larger than the 32\% found for the 3 GHz VLA-COSMOS survey \citep{delvecchio2017}, which is consistent with the fact that nearly one third of the radio counterparts we associate to X-ray sources lie below the theoretical $5\sigma_\text{rms}\simeq\,11.5\mu$Jy detection limit of that survey. 

Overall, we expect a false matching rate of $\lesssim3\%$ ($\sim40$ sources, which includes 4 sources flagged as `potentially spurious' in Section \ref{sec:radiodata}), based on the combined FMRs from the cross-matching of the individual catalogs, as well as a spurious fraction of $\lesssim2\%$. We hence deem our multi-wavelength catalog to be reliable.

\subsection{Redshifts of the Radio Sample}
\label{sec:specz}
Accurate redshift information for our radio sample is vital not only for the classification of AGN, but also for subsequent studies of the star-formation history of the universe. We therefore attempted to assign to each source its most reliable redshift through comparing the various redshifts (photometric or spectroscopic) we obtained from the different catalogs. First of all, we discarded all spectroscopic redshifts from the COSMOS master catalog (M. Salvato in prep.) that have a quality factor $Q_f < 3$, indicating an uncertain or poor spectroscopic redshift. All remaining sources for which a robust spectroscopic redshift is available are then assigned this value. The majority of sources additionally have photometrically determined redshifts available, from up to three different studies (e.g. \citealt{capak2007,laigle2016,jin2018}). We prioritize the photometric redshift from the Super-deblended catalog if available, as it is determined using prior photometric redshift information from other catalogs (e.g. COSMOS2015), but is re-computed with the inclusion of longer-wavelength data. However, any differences between these photometric redshifts are small by construction, as \citet{jin2018} force the Super-deblended redshift to be within 10\% of the prior value. If a Super-deblended redshift is unavailable, we instead use the photometric redshift from COSMOS2015 or the $i-$band selected catalog, in that order. We make an exception when the source is X-ray detected, in which case we assign it the photometric redshift from the \emph{Chandra} X-ray catalog \citep{marchesi2016}. These redshifts have been determined through SED fitting with the inclusion of AGN templates, and are therefore more appropriate for AGN, which form the bulk of our X-ray detected sample (Section \ref{sec:xray}). 

We compute the reliability of our redshifts, defined as $\sigma(z) = | z_\text{spec} - z_\text{phot} | / (1 + z_\text{spec})$ for the 584 sources which have both photometric and spectroscopic redshift information available. The normalized median absolute deviation \citep{hoaglin1983}, defined as $1.48$ times the median of $\sigma(z)$, is found to be $0.012$, indicating a very good overall consistency between the two redshifts. The fraction of sources with $\sigma(z) > 0.15$, the common threshold for defining `catastrophic failures', equals 4.8 per cent, with the main region of such failures being the optically fainter sources, as expected. We verified that the distribution of such failures in terms of $i-$band magnitude is similar to that in Figure 11 of \citet{laigle2016}, though with a slightly larger failure fraction at fainter magnitudes $i_\text{AB} \gtrsim 23$, which can be fully explained by our small sample size at these magnitudes.\footnote{While most of our photometric redshifts are from the Super-deblended catalog, and not from COSMOS2015, the former are by construction similar to the redshift adopted for the deblending prior, and as such comparing our photometric redshift accuracy with \citet{laigle2016} is justified.}

\begin{figure*}[!t]
	\centering
	\includegraphics[width=0.49\textwidth,height=5.0cm]{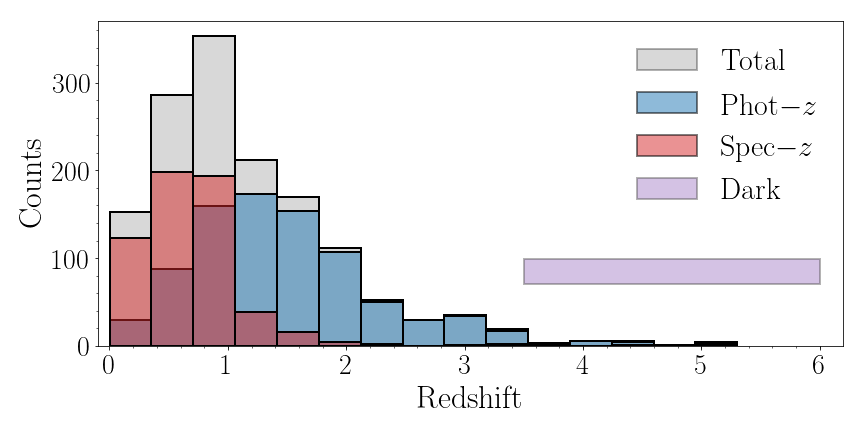} \hfill 
	\includegraphics[width=0.49\textwidth, height=5.1cm]{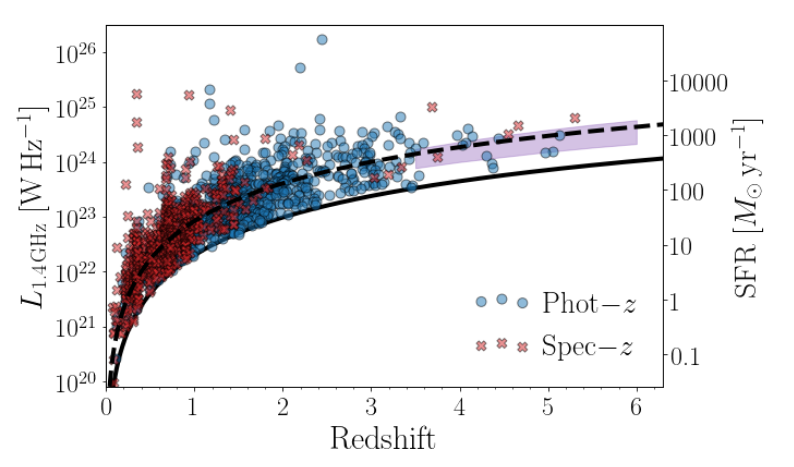}
	
	\caption{\textbf{Left:} Distribution of the radio-selected sample in redshift, for all 1437 sources with redshift information. Red and blue bars indicate spectroscopic and photometric redshifts respectively, and the grey histogram represents the full distribution. We additionally show the expected redshift range populated by the 29 `optically dark' sources -- sources without optical counterparts and redshift information, analyzed in Section \ref{sec:radioonly} -- via the purple bar. Out to $z\sim1$, nearly two-thirds of our redshifts are spectroscopic. \textbf{Right:} Rest-frame 1.4 GHz luminosity versus redshift for the COSMOS-XS 3 GHz-detected sample with reliable redshift information. The flux limit of the survey is indicated through the solid black line. Blue circles and red crosses indicate photometric and spectroscopic redshifts, respectively. The dashed, black line represents the flux limit of the 3 GHz VLA-COSMOS survey \citep{smolcic2017a} and is included for comparison. The COSMOS-XS survey constitutes a factor $\sim5$ increase in sensitivity compared to these wider radio data. We additionally show the redshift and luminosity range likely populated by the radio-detected sources without optical/near-IR photometry through the shaded purple area (Section \ref{sec:radioonly}). Star-formation rates (right ordinate axis) are computed assuming the radio emission is fully powered by star formation, and adopting a conversion that is independent of redshift, based on the local galaxy sample studied by \citet{bell2003}; see also Section \ref{sec:radioexcess}. Radio luminosities are converted to rest-frame 1.4 GHz using a fixed spectral index $\alpha=-0.70$. The theoretical detection limit (solid black line) is computed using $5$ times the RMS in the S-band image centre and scaled to 1.4 GHz rest-frame using an identical spectral index.}
	\label{fig:speczfrac_senscurv}
\end{figure*}

In summary, our sample contains 584 sources ($\sim41\%$) with spectroscopic and 853 sources with photometric redshifts (left panel of Figure \ref{fig:speczfrac_senscurv}). About two-thirds of our $z\lesssim1$ sample is detected spectroscopically, but the fraction of spectropscopic redshifts drops dramatically towards higher redshift. Additionally, a total of 103 sources have no redshift information. The majority of these (70 sources) were not cross-matched to any multi-wavelength counterpart. Only two sources without redshift information have optical/near-IR photometry, but nevertheless no robust photometric redshift could be obtained for these catalog entries. The median radio flux density (and bootstrapped uncertainty) of the sources without redshift information is $S_\text{3 GHz} = 10.5_{-1.2}^{+2.0}\,\mu$Jy, similar to the median of the full sample, which equals $S_\text{3 GHz} = 11.3_{-0.5}^{+0.4}\,\mu$Jy. The median flux density of the radio sources detected only in COSMOS-XS, i.e.\ without cross-matches to the Super-deblended or VLA-COSMOS catalogs, equals $7.2_{-0.7}^{+1.4}\,\mu$Jy -- below the formal detection limit of the latter survey, as expected.

We show the detection limit of the COSMOS-XS survey as function of redshift in the right panel of Figure \ref{fig:speczfrac_senscurv}. For ease of comparison with previous surveys, which have predominantly been performed at lower frequencies, we show the distribution of rest-frame 1.4 GHz luminosities, computed using a standard spectral index of $\alpha=-0.7$ even where multiple radio fluxes were available (see Section \ref{sec:radioexcess}). These luminosities are converted into star-formation rates adopting the conversion from \citet{bell2003} under the assumption that the radio emission is fully powered by star formation. In the following section, we will instead adopt a redshift-dependent conversion factor, as it has recently been shown to evolve with cosmic time \citep{magnelli2015,delhaize2017,calistrorivera2017}.

\begin{figure}[!t]
	\centering
	\includegraphics[width=0.45\textwidth]{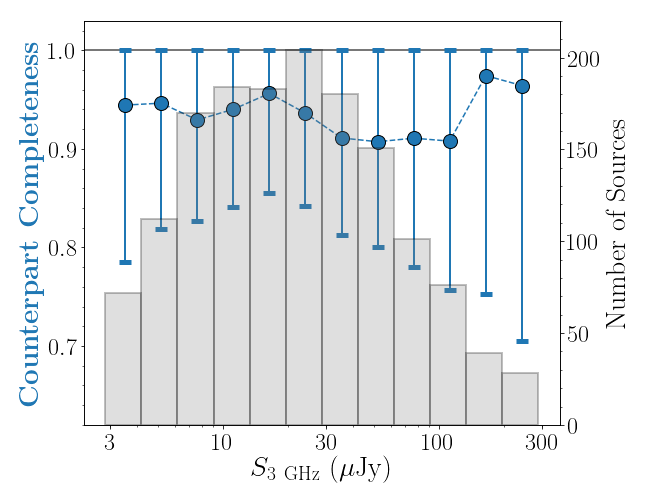}
	\caption{Counterpart completeness as a function of flux density for the S-band sample (left ordinate axis). Our S-band selected catalog contains counterparts with reliable redshifts for $93.3\%$ of all S-band detected sources, and all flux density bins are $\gtrsim90\%$ complete. No clear trend with flux density is seen, making it unlikely that a substantial number of low-SNR S-band sources are spurious, as for these no counterpart would be expected. The background histogram and corresponding right ordinate axis indicate the total number of sources in a given flux density bin.}
	\label{fig:counterpartcompleteness}
\end{figure}

Finally, in Figure \ref{fig:counterpartcompleteness}, we show the overall counterpart completeness of the 1540 S-band sources in twelve flux density bins. Uncertainties on the counting statistics were calculated following \citet{gehrels1986} for bins with fewer than 10 sources without optical counterparts, and were assumed to be Poissonian otherwise. We adopt these confidence limits for sparsely populated bins throughout this work. The completeness in all bins is upwards of 90\%, and no trend with radio flux density can be seen, indicating that the association of counterparts to our radio sources is not limited by the depth of the multi-wavelength photometry. Additionally, this indicates that it is unlikely that there are a substantial number of spurious radio sources within our 3 GHz image, as these would likely populate the low flux density bins and would typically not be associated to a multi-wavelength counterpart. Despite the lack of multi-wavelength counterparts for $\sim4.5\%$ of radio sources, a substantial fraction of these are simply faint at optical/near-infrared wavelengths, and as such are not spurious (see the discussion in Section \ref{sec:radioonly}).

\section{AGN IDENTIFICATION}
\label{sec:agn}

In this section, we outline the criteria used for identifying AGN among our radio-detected sources, making use of the wealth of data available over the COSMOS field. In the literature, numerous such multi-wavelength identifiers exist, based on both broad-band photometry and spectroscopy. In order to avoid potential biases due to incomplete coverage in various bands, we only utilize AGN diagnostics which are available for the vast majority of our radio-selected sample, which are outlined below. This excludes such diagnostics as inverted radio spectral indices (e.g.\ \citealt{nagar2000}), very long baseline interferometry (e.g.\ \citealt{herreraruiz2017}) and optical spectra (e.g.\ \citealt{baldwin1981}).\footnote{Radio spectral indices are available for 255 sources ($\sim18\%$ of the sample with redshift information). Eight of these have inverted spectral indices (7 from 1.4 - 3 GHz data, and a single source detected at 3 and 10 GHz), though seven were previously identified as AGN through other multi-wavelength diagnostics. The inclusion of an inverted spectral index as AGN diagnostic will therefore have a negligible effect on the overall AGN identification.}

In our panchromatic approach to AGN identification, we make use of SED fitting code {\sc{magphys}} \citep{dacunha2008,dacunha2015} to model the physical parameters of our galaxy sample. {\sc{magphys}} imposes an energy balance between the stellar emission and absorption by dust, and is therefore well-suited to model the dusty star-forming populations to which radio observations are sensitive. We stress that {\sc{magphys}} does not include AGN templates, and is therefore only suitable for modelling sources whose SEDs are dominated by star-formation related emission. However, we can use this to our advantage when identifying AGN based on their radio emission in Section \ref{sec:radioexcess}. Additionally, we use {\sc{AGNfitter}} \citep{calistrorivera2016,calistrorivera2017}, a different SED-fitting routine appropriate for AGN, in Section \ref{sec:sedfitting} to further mitigate this issue. We fit our full radio sample with {\sc{magphys}}, including all FUV to mm data. The radio data is not fitted, as an excess in radio emission is indicative of AGN activity and could therefore bias our results (Section \ref{sec:radioexcess}).

We will follow the terminology introduced in \cite{delvecchio2017} and \cite{smolcic2017b}, and divide the radio-detected AGN into two classes: the moderate-to-high luminosity AGN (HLAGN) and low-to-moderate luminosity AGN (MLAGN). These definitions refer to the radiative luminosity of AGN, resulting from accretion onto the supermassive black hole, which traces the overall accretion rate.\footnote{\citet{delvecchio2017} further argue that the HLAGN/MLAGN definitions closely resemble the widely used HERG/LERG nomenclature.} For efficiently accreting AGN ($\dot{m} \gtrsim 0.01\ \dot{m}_\text{Edd}$, where $\dot{m}$ is the mass accretion rate and $\dot{m}_\text{Edd}$ the Eddington rate), the bulk of the radiative luminosity is emitted by the accretion disk (UV) as well as in X-rays \citep{lusso2011,fanali2013}. Depending on both the orientation and the optical depth of the obscuring torus, this radiation may be (partially) attenuated and re-emitted in the mid-infrared (e.g.\ \citealt{ogle2006}). X-ray and MIR-based tracers of AGN activity therefore preferentially select high radiative luminosity AGN, and hence imply higher overall accretion rates. Conversely, low accretion rates ($\dot{m} \ll 0.01\ \dot{m}_\text{Edd}$) are associated with radiatively inefficient accretion, whereby the accretion disk is generally truncated and advection-dominated accretion takes over in the vicinity of the black hole (e.g.\ \citealt{heckmanbest2014}). As the timescale of such inflows is much shorter than the cooling time of the material, such ineffecient accretion produces little UV and X-ray emission. A recent study by \citet{delvecchio2018}, employing X-ray stacking on the 3 GHz-selected radio-excess AGN sample from VLA-COSMOS, indeed finds that below $z\lesssim2$ the accretion rates of such AGN are $\dot{m} \lesssim 0.01\ \dot{m}_\text{Edd}$, with only 16\% of this sample being individually X-ray detected, implying overall inefficient accretion for the typical radio-excess AGN. They additionally do not find any correlation between AGN X-ray and radio luminosity at a fixed redshift. Instead, the identification of AGN with lower accretion rates and radiative luminosities, referred to here as MLAGN, thus relies predominantly on radio-based diagnostics. These are effectively based on the fact that, for such AGN, the multi-wavelength star-formation rate indicators are discrepant, as will be clarified in Section \ref{sec:radioexcess}. It must be noted, however, that a hard division between high- and moderate-luminosity AGN does not exist, and we therefore follow \citet{delvecchio2017} by applying the tags `moderate-to-high' and `low-to-moderate' to indicate the overlap between the classes. This further serves to illustrate that there is no one-to-one relation between the class an AGN belongs to and its accretion rate. We will study the various sets of AGN in more detail in a future paper, and instead focus on the classification in this work.

\subsection{HLAGN}
\label{sec:xray}
In the context of this paper, we identify a source as an HLAGN if it satisfies any of the following criteria:
\begin{itemize}
	\item The source shows a $\geq2\sigma$ excess in X-ray luminosity compared to its FIR-derived star-formation rate, based on the relations from \citet{symeonidis2014}.
	\item The source exhibits mid-IR IRAC colors that place it within the \citet{donley2012} wedge, provided it lies at $z\leq2.7$.
	\item The source shows a significant AGN-component in the form of a dusty torus or accretion disk, based on SED fitting.
\end{itemize}
We expand on each of these criteria in the following subsections.

\subsubsection{X-ray AGN}
A subset of HLAGN are characterized by a high X-ray luminosity, which is thought to originate from the accretion disk around the central supermassive black hole (SMBH). In the conventional picture \citep{heckmanbest2014}, this disk is surrounded by a hot corona, which boosts the energy of the seed photons from the accretion process through inverse Compton scattering into the X-ray regime. The accretion disk is further thought to be obscured by a dusty torus, which -- if sufficiently dense -- may absorb even the hard X-rays produced by the AGN. Nevertheless, in the scenario of low obscuration, AGN-powered X-ray emission can be orders of magnitude brighter than X-rays expected from star-formation related processes, which arise primarily from high-mass X-ray binaries \citep{fabbiano2006}. In order to classify our X-ray detected sources as AGN, we make use of the X-ray properties of star-forming galaxies derived by \citet{symeonidis2014}. They found that typical SFGs have a relation between their FIR-luminosity and soft band ([$0.5-2]$ keV) X-ray luminosity given by $\log L_\text{FIR} = \log L_\text{[0.5-2] keV} + 4.55$, with a $2\sigma$ scatter around this relation of $0.74$ dex. We classify sources with an X-ray excess above this $2\sigma$ scatter as HLAGN. 

For the AGN classification we extract $[0.5 - 10]$ keV obscuration-corrected X-ray luminosities from the \emph{Chandra} COSMOS Legacy catalog presented in \citet{marchesi2016}.\footnote{In case an X-ray source was assigned a different redshift than given in the \emph{Chandra} catalog, we re-computed its X-ray luminosity using the updated value.} We scale X-ray luminosities between different bands with a power law index of $\Gamma = 1.4$, defined such that the X-ray luminosity follows $L_X \propto \nu^{1-\Gamma}$. In the following, we will quote X-ray luminosities in the $[0.5-8]$ keV range. Out of the 108 cross-matches we obtained within $1\farcs4$ with this catalog, we identify 106 sources as X-ray AGN.\footnote{Upon adopting a steeper X-ray photon index of $\Gamma = 1.7$, as may be more appropriate for star-forming galaxies (e.g.\ \citealt{lehmer2010}), we instead identify one additional X-ray AGN. However, including this additional source as an X-ray AGN has no impact on our overall conclusions.} Had we adopted a fixed X-ray luminosity threshold of $L_\text{X} = 10^{42} \ \text{erg s}^{-1}$, as is common in the literature (e.g. \citealt{wang2013,delvecchio2017,smolcic2017b}), we would have missed an additional 7 X-ray sources at low redshift that are only modestly X-ray luminous, but are nonetheless substantially offset from the relations from \citet{symeonidis2014}. Our main motivation for adopting a threshold dependent on $L_\text{FIR}$ is however to avoid the misclassification of highly starbursting sources, as SFRs upwards of $\sim300\ M_\odot\ \text{yr}^{-1}$ are expected to generate  X-ray luminosities in excess of $L_\text{X} \sim 10^{42} \ \text{erg s}^{-1}$. We therefore regard a selection based on the comparison of FIR- and X-ray luminosities to be more robust in general. Based on the discussion in Appendix \ref{app:hlagn}, where we employ X-ray stacking, we further conclude that we are minimally affected by incompleteness issues, which may arise from the relatively shallow X-ray data, resulting in rather unconstraining upper limits on the X-ray luminosities of the typical high-redshift ($z\gtrsim2$) source.

\begin{figure}[tb]
	\centering
	\includegraphics[width=0.49\textwidth]{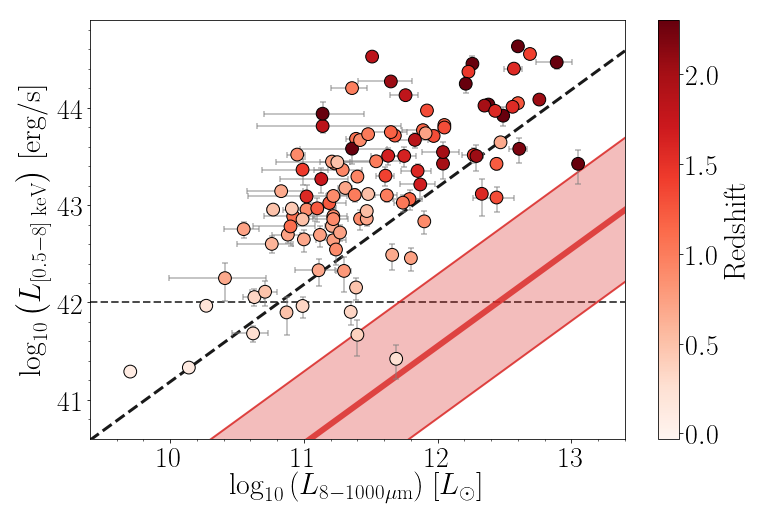}
	\caption{X-ray luminosity in the $[0.5 - 8]$ keV energy band versus total infrared ($8-1000\mu$m) luminosity for our X-ray detected sample, color-coded by redshift. The solid diagonal line represents the X-ray versus total-infrared luminosity relations for star-forming sources from \citet{symeonidis2014}, scaled to a full $[0.5 - 8]$ keV using a power law exponent $\Gamma=1.4$. The red, shaded region indicates the $2\sigma$ scatter around this relation (equivalent to 0.74 dex). All soures that fall outside of this scatter are identified as X-ray AGN in this work. The diagonal, black line is shown for comparison and represents the AGN luminosity threshold from \citet{alexander2005}, which is more conservative in identifying X-ray AGN than the threshold we adopt here. The dashed, horizontal line indicates a constant X-ray luminosity threshold of $10^{42}\ \text{erg s}^{-1}$, which is a typically used threshold for identifying AGN in the literature at low redshift. All but two of our X-ray detected sources are substantially offset from the \citet{symeonidis2014} relations, including 7 low-redshift sources with $L_X \lesssim 10^{42}$ erg/s.}
	\label{fig:xraysfr}
\end{figure}

\subsubsection{MIR AGN}

Sources that fall within the class of high-luminosity AGN are believed to be surrounded by a warm, dusty torus, which will absorb and re-radiate emission emanating from the region around the central SMBH. This gives rise to a specific MIR-continuum signature associated to predominantly dusty and obscured AGN. Early work in the identification of AGN based on MIR-colors was done by \citet{lacy2004}, based mostly on the \emph{Spitzer}/IRAC colors of local Seyfert galaxies. Due to intrinsic reddening of high-redshift sources, these criteria are not optimized for galaxies at moderate redshift ($z\gtrsim0.5$), and we therefore use the adapted criteria from \cite{donley2012} to identify obscured HLAGN. We locate sources within the \citet{donley2012} wedge, defined through their equations (1) and (2) and identify such sources as MIR AGN. As the MIR-colors of dusty star-forming galaxies at high redshift ($z\gtrsim3$) closely resemble those of obscured AGN (e.g.\ \citealt{stach2019}), we restrict our analysis to $z\leq2.7$, as \citet{donley2012} are increasingly biased above this redshift. We note these MIR-criteria are somewhat conservative in order to minimize the occurrence of false positives, and the MIR-identification becomes less complete for X-ray faint AGN. Overall, we recover 28 AGN based on their \emph{Spitzer}/IRAC colors. While only $\sim60\%$ of our sample has reliable IRAC photometry in all four channels, and hence we cannot robustly place the remaining sources within the \citet{donley2012} wedge, this incompleteness has negligible effect on our overall AGN identification (Appendix \ref{app:hlagn}).

\subsubsection{AGN SED-fitting}
\label{sec:sedfitting}
HLAGN are expected to show a composite multi-wavelength SED, exhibiting signs of both star-formation and AGN-related processes. A spectral decomposition will therefore detail the relative contribution of these two components, allowing AGN to be identified as such if their emission dominates over the contribution from star formation at certain wavelengths. We use {\sc{AGNfitter}} \citep{calistrorivera2016} to fit the far-UV to FIR SEDs of our radio-selected sample. {\sc{AGNfitter}} is a publicly available python-based SED-fitting algorithm implementing a Bayesian Markov Chain Monte Carlo method to fit templates of star-forming and AGN components to observed multi-wavelength galaxy photometry. Two such AGN components are fitted: an accretion disk, which predominantly emits at UV- and optical wavelengths, and a warm, dusty torus that contributes mostly to the MIR-continuum. The SED-fitting further includes UV/optical emission from direct starlight, as well as dust-attenuated stellar emission in the infrared.

As {\sc{AGNfitter}} utilizes a Monte Carlo method in its SED-fitting procedure, its output includes realistic uncertainties on any of its computed parameters, such as the integrated luminosities in the various stellar and AGN components. These uncertainties are particularly informative for galaxies with no or little FIR-photometry, as in this case the long-wavelength SED is largely unconstrained. This is in contrast to SED fitting codes that impose energy balance between the stellar and dust components, such as {\sc{magphys}} \citep{dacunha2008,dacunha2015} and {\sc{sed3fit}} \citep{berta2013} which is built upon the former and extended to include AGN templates. We opt for a Bayesian algorithm without energy balance, as it has been shown that dust and stellar emission can be spatially offset in high-redshift dusty galaxies such that imposing energy balance may be inaccurate (e.g. \citealt{hodge2016}). In addition, it allows for the comparison of realistic probability distributions for the integrated luminosities of the various galaxy and AGN components, enabling us to separate the populations based on physical properties, rather than based on the goodness of fit. We compare our results with those obtained with {\sc{sed3fit}} by \citet{smolcic2017b} in Appendix \ref{app:sedfitting}.

Prior to the fitting, we account for uncertain photometric zeropoint offsets and further potential systematic uncertainties by adding a relative error of $5\%$ in quadrature to the original error on all photometric bands between $U$ and MIPS 24$\,\mu$m, similar to e.g.\ \citet{battisti2019}. This further serves to guide the fitting process into better constraining the spectrum at FIR wavelengths, where photometric uncertainties are generally large. Without such an adjustment, the fitting would be dominated by the small uncertainties on the short-wavelength photometry, and occasionally fail to accurately model the FIR component. 

We then identify AGN via a comparison of the integrated luminosities in both the torus and accretion disk components with, respectively, the stellar-heated dust continuum and the direct optical and near-UV stellar light, taking into account the probability distributions of these integrated luminosities. This comparison then directly takes into account the reliability of the photometry, as large photometric uncertainties will naturally lead to a broad probability distribution in the integrated luminosities of the various components. Therefore, this procedure only includes AGN that can reliably be identified as such, similar to e.g.\ \citet{delvecchio2014,delvecchio2017}, who compare the best-fitting SEDs with and without AGN templates, and require the former to be a better fit at the $99\%$ confidence level in order to identify it as AGN. We slightly modify this procedure for sources without any FIR-photometry, and expand on the exact criteria we employ in Appendix \ref{app:agnfitter}. Overall, we identify 149 sources as HLAGN based on SED fitting, with 51 (78) being identified solely through a MIR-torus (accretion disk) component. A further 20 sources are classified as an AGN based on both of these features.

\begin{figure}[h]
	\centering
	\includegraphics[width=0.35\textwidth]{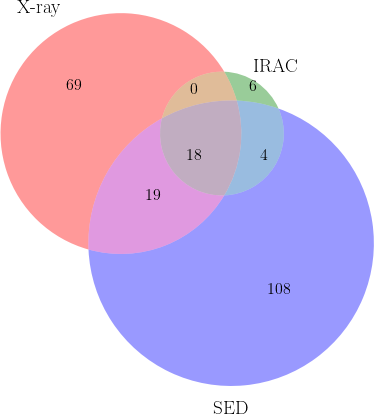}
	\caption{Venn diagram showing the overlap between the various AGN diagnostics used for identifying HLAGN. The areas of circles and overlapping regions are roughly proportional to the number of sources within this category.}
	\label{fig:venn}
\end{figure}

We show the overlap between the three different methods utilized for identifying HLAGN in Figure \ref{fig:venn}. As expected, the subset of AGN identified through mid-IR colors largely overlaps with those found through SED-fitting, such that the X-ray and SED-fitted AGN make up the bulk of the total set of HLAGN.

\subsection{MLAGN}\label{sec:radioexcess}
Whereas AGN selected through X-ray emission, MIR-colors and SED-fitting diagnostics preferentially identify HLAGN, which form the subset of AGN powered by efficient accretion, a second, low-luminosity AGN population is most readily detected through its radio properties. We assign sources to the class of low-to-moderate luminosity AGN (MLAGN) if they are not identified as HLAGN and satisfy one of the following criteria:
\begin{itemize}
	\item The source exhibits radio-emission that exceeds (at the $2.5\sigma$ level) what is expected from star formation, based on the radio-FIR correlation.
	\item The source exhibits red rest-frame $\left[\text{NUV}-r^+\right]$ colors, corrected for dust attenuation, typically indicating a lack of star formation.
\end{itemize}

\subsubsection{Radio-excess AGN}
The FIR-radio correlation describes a tight interconnection between the dust luminosity of a star-forming galaxy and its low-frequency radio luminosity. This connection arises because the same population of massive stars that heats up dust, causing it to re-radiate its energy in the FIR, produces supernovae that generate relativistic particles emitting synchrotron radiation at radio frequencies. However, galaxies that host an AGN may have their radio emission dominated instead by the active nucleus, and will therefore be offset from the FIRC. To quantify this, we define the correlation parameter $q_\text{TIR}$ as the logarithmic ratio of a galaxy's total IR-luminosity $L_\text{TIR}$, measured between (rest-frame) $8-1000\,\mu$m, and its monochromatic radio luminosity at rest-frame 1.4 GHz, $L_{1.4\text{GHz}}$ (following e.g. \citealt{bell2003,thomson2014,magnelli2015,delhaize2017,calistrorivera2017}):

\begin{align}
	q_\text{TIR} = \log_{10}\left(\frac{L_\text{TIR}}{3.75\times 10^{12} \text{ W}}\right) - \log_{10} \left(\frac{L_{1.4\text{GHz}}}{\text{W/Hz}}\right) \ .
	\label{eq:qtir}
\end{align}

\begin{figure*}[t]
	\centering
	\hspace*{-0.5cm}
	\includegraphics[width=\textwidth]{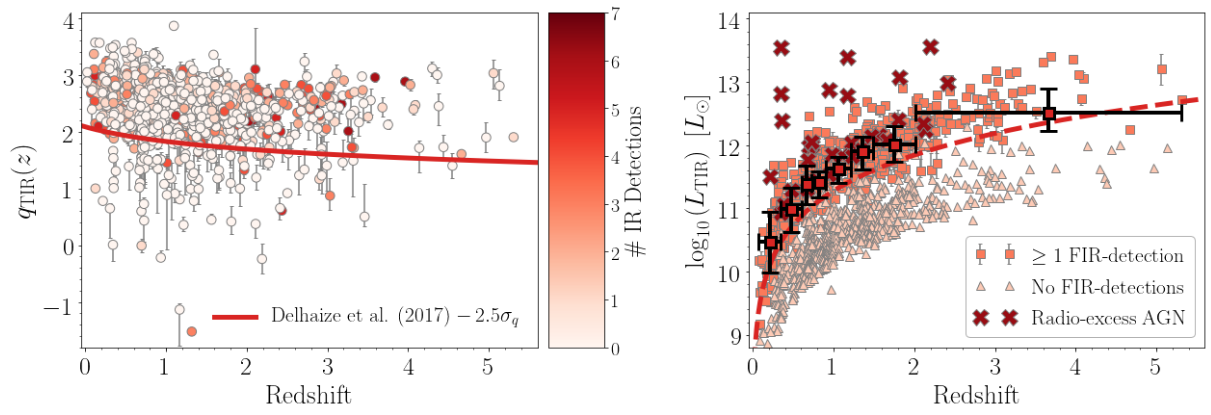}
	\caption{\textbf{Left:} FIRC-parameter $q_\text{TIR}$ as a function of redshift for the S-band detected sample of 1437 sources. Sources are colored by the number of detections in the FIR (including \emph{Herschel}/PACS \& SPIRE, as well as SCUBA2, JCMT/AzTEC and IRAM/MAMBO). The solid red line constitutes the redshift-dependent trend of $q_\text{TIR}(z)$ as determined for star-forming galaxies by \citet{delhaize2017}, minus $2.5\times$ the intrinsic scatter of $0.31\,$dex, which constitutes our redshift-dependent threshold for radio excess. The 110 sources below this line are then identified as radio-excess AGN. \textbf{Right:} Empirically determined sensitivity curve of our \emph{Herschel} observations, showing the redshift-dependent FIR-luminosity of our sample, either computed through {\sc{magphys}} (orange squares) or via the FIRC, assuming a conservative $q_\text{TIR}(z)$ (triangles and crosses). The sources with good FIR-photometry are binned in redshift (large red squares). The empirical detection limit is then determined via a fit through the lower errorbars, and is shown via the dashed red line. Radio sources without far-infrared photometry that fall above this detection threshold are identified as AGN. This diagnostic allows us to quantify radio excess for sources lacking FIR-photometry, and identifies a total of 62 radio-excess AGN.}
	\label{fig:radioexcess}
\end{figure*}

The factor $3.75\times 10^{12}$ is the central frequency of the total-IR continuum ($8-1000\,\mu$m) in Hz and serves as the normalization. There is now a growing consensus that  $q_\text{TIR}$ is a function of redshift \citep{magnelli2015,delhaize2017,calistrorivera2017}, for reasons that are still rather poorly understood. Nevertheless, we utilize a redshift-dependent threshold in terms of $q_\text{TIR}$ to identify galaxies with radio excess based on what is expected from the FIRC. We show the distribution of $q_\text{TIR}$ as a function of redshift for our sample of radio-detected sources in Figure \ref{fig:radioexcess}, with the FIR-luminosities obtained from {\sc{magphys}}.\footnote{Using energy-balance is appropriate here, as it allows us to associate FIR-luminosities even to sources without good photometric coverage at these wavelengths. Adopting a code without energy balance would instead result in artificially low FIR-luminosities and as such biases sources towards being radio-excess AGN. We study the effect of incompleteness in our FIR-photometry and the assumption of energy balance in Appendix \ref{app:firphot}.} Rest-frame 1.4 GHz luminosities are determined using the measured spectral index for the required $K$-corrections if available. When only a single radio flux is available, a spectral index of $\alpha=-0.7$ is assumed instead. The luminosities are then calculated through 

\begin{align}
    L_{1.4\text{ GHz}} = \frac{4\pi D_L^2 }{(1 + z)^{1+\alpha}} \left( \frac{1.4\text{ GHz}}{3\text{ GHz}}\right)^\alpha S_{3\text{ GHz}}  \ .
\end{align}

Here $D_L$ is the luminosity distance at redshift $z$ and $S_\text{3 GHz}$ is the observed flux density at 3 GHz. The uncertainty on the luminosity is computed by propagating the error on the flux density and -- if the source is detected at $\geq2$ radio frequencies -- the spectral index, i.e. the source redshift is taken to be fixed. The error on $q_\text{TIR}$ further includes the propagated uncertainty on the FIR-luminosity returned by {\sc{magphys}}.

To quantify radio excess, we adopt the redshift-dependent $q_\text{TIR}$ determined for star-forming galaxies by \citet{delhaize2017}. They determine a best-fitting trend of $q_\text{TIR,D17} = 2.86 \times (1 + z)^{-0.19}$, with an intrinsic scatter around the correlation of $\sigma_q = 0.31$. Their best fit takes into account the sample selection at both radio and far-infrared wavelengths through a two-sided survival analysis. As such, \citet{delhaize2017} take into account that radio-faint star-forming galaxies -- those with a value of $q_\text{TIR}$ above the typical correlation -- are preferentially missed in radio-selected samples, in particular at high-redshift due to the negative radio $K-$correction. We therefore adopt their median redshift-dependent value for $q_\text{TIR}$ appropriate for star-forming galaxies, minus $2.5\times\sigma_q$, as our threshold for identifying radio-excess AGN. That is, our threshold identifies a radio source as an AGN if it lies below the median far-infrared-radio correlation for star-forming galaxies at more than $2.5\sigma$ significance. Adopting the \citet{delhaize2017} results appropriate for star-forming sources, and taking into account the intrinsic scatter about the correlation, minimizes the effect of selection biases and incompleteness in our multi-wavelength photometry on the AGN classification. Overall, we identify a total of $110$ radio-excess AGN via this method.

However, this number of AGN may be affected by the fact that we do not have radio spectral indices for $\sim80\%$ of our sample (see also \citealt{gim2019}). To test the effect of assuming a fixed spectral index for these sources, we re-compute the number of radio-excess AGN by assigning every source detected solely at 3 GHz a spectral index drawn from a normal distribution, centered around $\alpha=-0.70$, with a standard deviation of $\sigma=0.30$ (similar to the intrinsic scatter in the radio spectral indices found by \citealt{smolcic2017a} in the 3 GHz VLA-COSMOS survey). We then run this procedure $200$ times and find the mean number of radio-excess AGN to be $N_\text{AGN} = 120$, with a standard deviation of $5$ sources. It is unsurprising that the typical number of radio-excess AGN increases slightly when a distribution of spectral indices is assumed, as the number of star-forming sources is $\sim12\times$ greater than the number of AGN. As such, it is more likely for a galaxy classified as star-forming when $\alpha=-0.70$ is assumed to scatter below the AGN threshold than for an AGN to scatter into the star-forming regime. However, the minor increase of $\sim10$ radio-excess AGN we find when adopting such a distribution of spectral indices does not change our conclusions that radio-excess AGN make up only a small fraction of the $\mu$Jy radio population.

While the radio-excess criterion described above constitutes a clear way to identify AGN for sources with well-constrained FIR-luminosities, only 50\% of our sample is detected in the far-infrared at $\geq 3\sigma$. To improve the completeness of our sample of radio-excess AGN, we utilize the distribution of FIR-luminosities for the sources with at least one $3\sigma$ detection at any of the FIR-wavelengths, and compare this with expected FIR-luminosities of the \emph{Herschel}-undetected sources. For each of the latter, we compute the FIR-luminosity through the far-infrared radio correlation, assuming the previously mentioned relation for star-forming sources by \cite{delhaize2017}, again adopting their normalization minus $2.5$ times their intrinsic scatter about the correlation. Effectively, we thus compute a conservative FIR-luminosity of our \emph{Herschel}-undetected sample at the $2.5\sigma$ level, assuming all radio emission is powered by star formation. We compare the FIR-luminosities derived in this manner with the distribution of luminosities for our sources with well-constrained FIR photometry in the lower panel of Figure \ref{fig:radioexcess}. We fit a power law through the 16$^\text{th}$ percentile of the distribution of $\log L_\text{TIR}$ in each redshift bin for sources with FIR-detections, and thus empirically determine the detection threshold of sources with a given dust luminosity. Sources which fall above the median FIR-luminosity determined for the sample with \emph{Herschel}-detections, yet are themselves undetected in the FIR, are also identified as `inverse' radio-excess AGN. This constitutes a total of 62 sources, shown via the red crosses in the lower panel of Figure \ref{fig:radioexcess}. A substantial number of these, 46 in total, were previously identified through the normal radio-excess criterion. This substantiates that the energy balance {\sc{magphys}} applies to determine far-infrared luminosities is typically a good assumption for these sources (see also \citealt{dudzeviciute2019}).

\subsubsection{Red, quiescent AGN}
The class of MLAGN can be further extended by including red galaxies, as $-$ once obscuration by dust has been corrected for $-$ such colors indicate a cessation of star formation. We quantify this through the rest-frame $[\text{NUV} - r^+]$-colors of our sources, which we model through integrating the best-fitting, unattenuated {\sc{magphys}} SED over the GALEX NUV $2300$ \AA\ and Subaru $r^+$-band filters. We follow \citet{ilbert2010} and define sources with $[\text{NUV}-r^+] > 3.5$ as quiescent galaxies, but limit this analysis to sources at $z\leq2$, as we do not accurately measure the rest-frame UV emission for sources at high redshift. As radio emission traces star formation, and quiescent sources by definition lack significant star-formation activity, we identify radio sources with red rest-frame colors as MLAGN. We find 50 such sources and note that $56\%$ of these were already previously identified through the (inverse) radio excess criterion, similar to what is found by \cite{smolcic2017b}. We verified that there is no trend between the NUV/optical colors and redshift, as such a trend may be indicative of an inaccurate extrapolation of a galaxy's SED by {\sc{magphys}} to rest-frame NUV wavelengths, resulting in the misclassification of red, quiescent MLAGN.

\begin{figure*}[t]
	\centering
	\includegraphics[width=\textwidth]{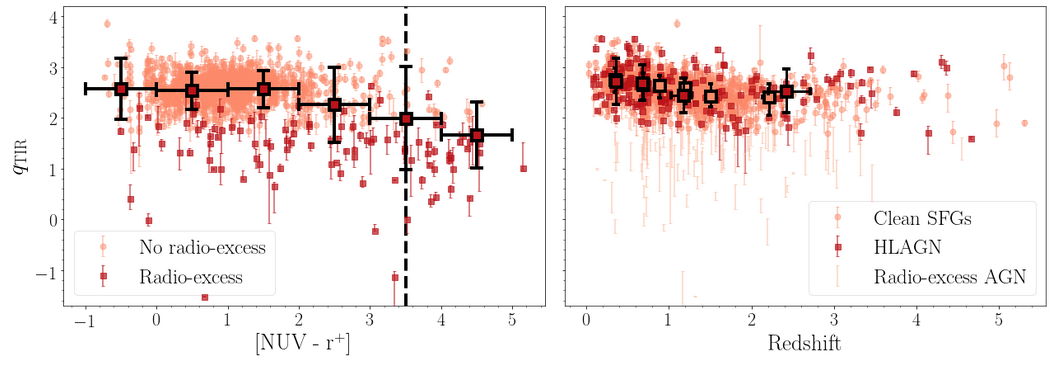}
	\caption{\textbf{Left:} The radio-FIR correlation parameter $q_\text{TIR}$ (Equation \ref{eq:qtir}) as a function of rest-frame $[\text{NUV} - r^+]$-colors for all sources at $z\leq2$. The sample is divided into radio-excess AGN (red squares) and sources primarily powered by star formation (orange circles). The dashed black line represents the threshold for identifying sources as red, quiescent MLAGN. The large squares indicate the mean values per bin, and the binwidth is shown through the horizontal errorbars. The vertical errorbars represent the $1\sigma$ standard deviation per bin. On average, redder sources show stronger radio emission for a given FIR-luminosity, which is consistent with a substantial fraction of these sources being radio-excess AGN. \textbf{Right:} $q_\text{TIR}$ as a function of redshift for the clean star-forming sample (orange circles) and HLAGN without radio-excess (red squares). The large orange and red squares represent the binned values for the clean SFGs and HLAGN, respectively. No significant difference is present between the distribution of $q_\text{TIR}$ for the two samples, indicating that AGN without radio-excess have radio properties consistent with arising solely from star formation.}
	\label{fig:radioexcessperagn}
\end{figure*}

We show the relation between the FIRC-parameter $q_\text{TIR}$ and rest-frame $[\text{NUV} - r^+]$-colors for sources at $z\leq2$ in Figure \ref{fig:radioexcessperagn} (left panel). On average, redder sources exhibit lower values of $q_\text{TIR}$, and hence constitute a higher fraction of radio-excess AGN. A total of 22 sources $-$ nearly half of the sources with $[\text{NUV}-r^+] > 3.5$ $-$ nevertheless show radio emission consistent with originating solely from star formation, which may indicate that these sources are in fact low-level (dusty) SFGs without substantial AGN activity in the radio (though five are identified as HLAGN instead). The determination of whether these objects have an AGN contribution is however complicated by the fact that only four of these sources have detections in the far-infrared, such that the modelled dust continuum emission of these objects is determined solely through the energy balance that {\sc{magphys}} imposes. This adds an additional layer of uncertainty to their distribution of $q_\text{TIR}$. For this reason, as well as the general observation that red, early-type galaxies are typically linked with radio-bright AGN hosts (e.g. \citealt{rovilos2007,smolcic2009,cardamone2010,delvecchio2017}), we identify these sources as quiescent (ML)AGN nevertheless. Ultimately, the sample of red rest-frame optical/NUV sources not identified as AGN through other means only concerns a small number of sources (only $1.5\%$ of our radio sample, or $6.6\%$ of all AGN), and their inclusion has a negligible effect on the number counts we derive in Section \ref{sec:composition}, with all results being consistent within the uncertainties if we include these sources in the clean SFG sample instead. In fact, omitting these sources from the sample of MLAGN further strengthens our conclusions that AGN make up only a small fraction of the $\mu$Jy radio population.

\subsection{AGN Identification Summary}
\label{sec:agnsummary}
\begin{figure}[h]
	\centering
	\includegraphics[width=0.4\textwidth]{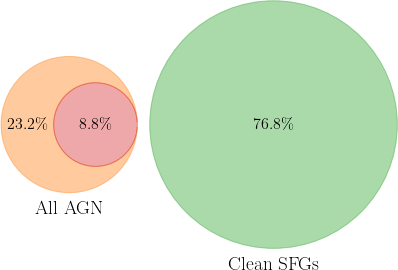}
	\caption{Visual representation of the different radio populations examined in this paper. The subset of all AGN is shown in orange, and combines both MLAGN and HLAGN. AGN exhibiting a radio excess are shown in red, and form less than half of the total AGN population. Clean star-forming sources are shown in green, and form the remainder of the radio sample.}
	\label{fig:vennagn2}
\end{figure}

The results of the AGN identification process are listed in Table \ref{tab:agn}, and are additionally summarized as a Venn diagram in Figure \ref{fig:vennagn2}. We find a total of 334 AGN in our sample ($23.2\pm1.3\%$, where the error represents the $1\sigma$ Poissonian uncertainty), using the five different diagnostics detailed in the previous sections. Combined, our AGN sample contains 224 HLAGN and 110 MLAGN. Overall, $\sim64\%$ of the sample was identified using just a single AGN tracer, whereas the remaining AGN were identified as such with up to four diagnostics. This exemplifies the importance of using various tracers of AGN activity, as the different diagnostics trace intrinsically different populations.

Less than half of the AGN we identify show an excess in radio emission with respect to the radio-FIR correlation, as is shown in the Venn diagram in Figure \ref{fig:vennagn2} by the red circle. We overplot all AGN without radio-excess on the radio-FIR correlation in Figure \ref{fig:radioexcessperagn} (right panel), which shows that these sources have radio emission that is fully consistent with originating from star formation. Therefore, only $8.8\pm0.8\%$ of our radio sample has radio emission that is clearly not from a star-forming origin.

\begin{deluxetable}{lllll}
	\tabletypesize{\footnotesize}
	\tablecolumns{5}
	\tablewidth{0.5\textwidth}
	\tablecaption{Summary of the AGN identification.}
	
	\tablehead{
		\colhead{\textbf{Method}} &
		\colhead{\textbf{HLAGN}} &
		\colhead{\textbf{Fraction}} &
		\colhead{\textbf{MLAGN}\tablenotemark{a}} &
		\colhead{\textbf{\%age}}
	}
	
	\startdata
	
	X-ray & 106 & 7.4\% & - & -  \\
	IRAC & 28 & 1.9\% & - & - \\
	SED-fitting & 149 & 10.4\% & - & - \\
	\hspace*{6pt} $\bullet$ Torus & 71 & 4.9\% & - & -\\
	\hspace*{6pt} $\bullet$ Disk & 98 & 6.8\%  & - & -\\
	Radio-excess & 25 & 1.7\% & 85 & 5.9\% \\
	Inverse radio-excess & 19 & 1.3\% & 43 & 3.0\% \\
	$[\text{NUV} - r^+]$ & 5 & 0.3\% & 45 & 3.1\% \\
	\tableline 
	\textbf{Total}\tablenotemark{b} & 224 & 15.6\% & 110 & 7.7\%
	\enddata
	
	\tablenotetext{a}{MLAGN are, by definition, not identified through the X-ray, IRAC or SED-fitting criteria.}
	\tablenotetext{b}{The total does not equal the sum of all rows, as a single AGN may be identified through multiple diagnostics.}
	\label{tab:agn}
\end{deluxetable}


\section{COMPOSITION OF THE ULTRA-FAINT RADIO POPULATION}
\label{sec:composition}

\subsection{The Ultra-faint Radio Population}
\label{sec:countsvsflux}
\begin{figure*}[!ht]
	\centering
	\includegraphics[width=\textwidth]{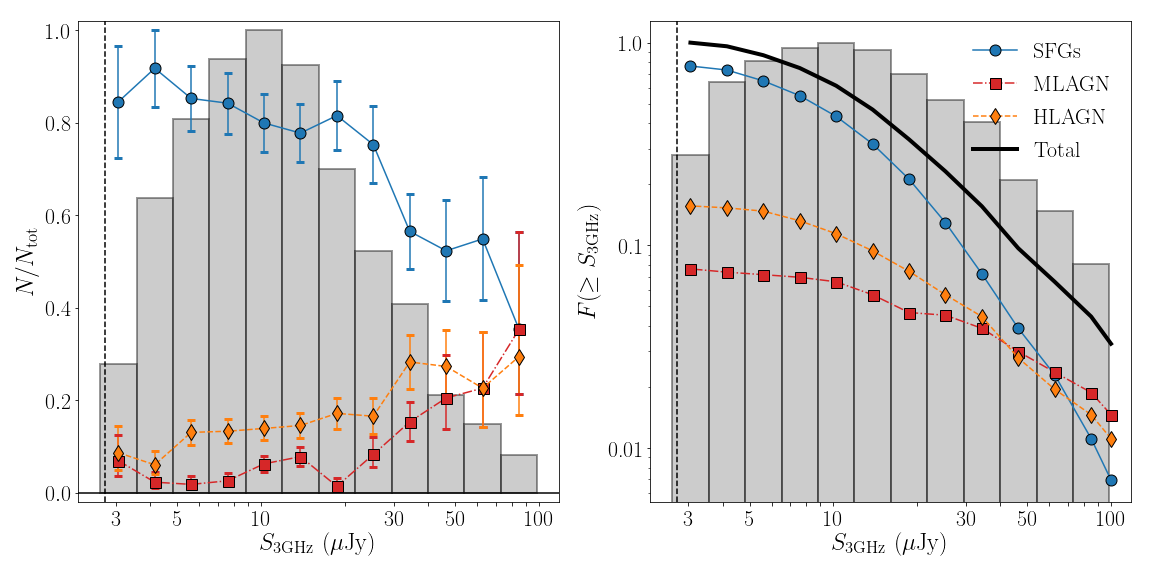}	
	\caption{\textbf{Left:} the fraction of sources of a given type (SFGs, MLAGN and HLAGN) as a function of flux density. Errorbars represent the Poissonian uncertainties for $>10$ sources per bin, or the confidence limits from \citet{gehrels1986} otherwise. In both panels, the vertical dashed grey line indicates the $5\sigma$ detection limit of the COSMOS-XS survey, and the grey histogram shows the normalized number of sources in a given logarithmic 3 GHz flux density bin. \textbf{Right:} the cumulative fraction of sources as a function of flux density, defined to increase towards lower flux densities. Overall, $\sim75\%$ of sources make up the sample of clean, star-forming galaxies, while below $\sim30\,\mu$Jy, the fraction of such SFGs increases to $\gtrsim80\%$.}
	\label{fig:agnwithflux}
\end{figure*}

In Figure \ref{fig:agnwithflux}, we show both the fractional and cumulative contribution of the different radio populations as a function of 3 GHz flux density. We restrict our analysis to sources with flux densities $<100\,\mu$Jy, which constitute $\sim98\%$ of our sample, because of poor statistics at our bright end. At relatively high flux densities between $50 - 100\,\mu$Jy, the radio population remains fairly equally split among the combination of MLAGN and HLAGN and clean star-forming galaxies, though our modest sample size at these fluxes results in significant uncertainties. Nevertheless, the class of MLAGN dominates the population of AGN at $S_\nu \gtrsim 50-100\,\mu$Jy, which is unsurprising as the bulk of this population is made up of sources that show radio excess, and are therefore radio-bright by definition. At flux densities $\lesssim30\,\mu$Jy, which constitutes 86\% percent of our sample, we observe a clear increase in star-forming sources, reaching a fractional contribution of $\gtrsim80\%$ in the lowest flux density bins. Cumulatively, our sample reaches $50\%$ star-forming sources at flux densities $\sim10\mu$Jy, and overall is made up for $\sim75\%$ by sources with no hints of AGN activity. At our detection limit of $\sim2.7\mu$Jy, approximately $85\%$ of the sample is made up of clean SFGs.

\begin{figure*}[!ht]
	\centering
	\includegraphics[width=\textwidth]{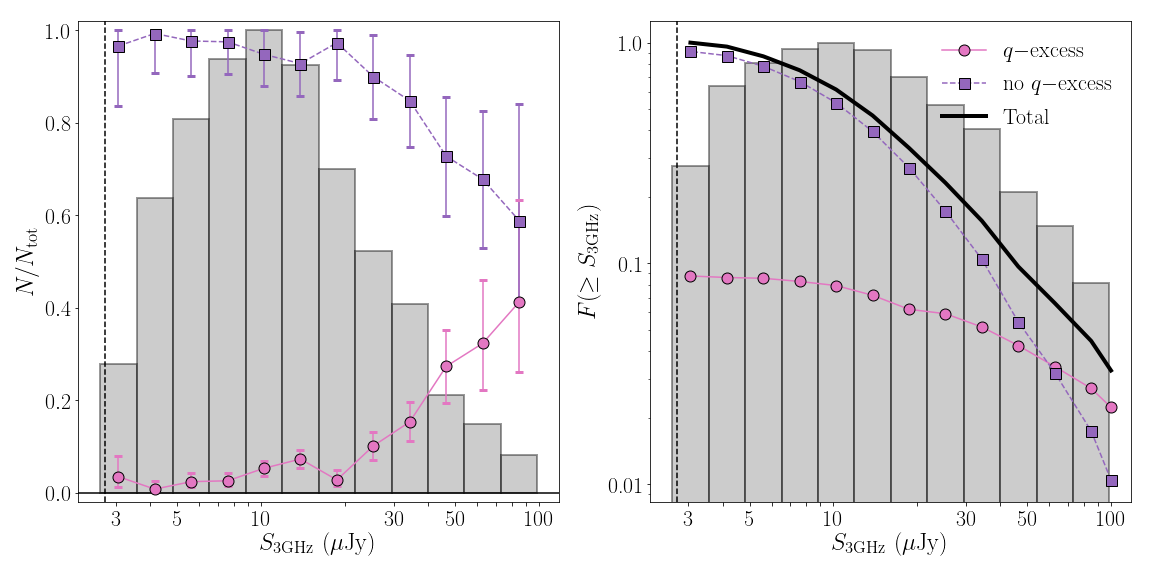}
	\caption{\textbf{Left:} the fraction of sources with (without) radio excess in pink (purple) as a function of 3 GHz flux density. The datapoints are shown with Poissonian uncertainties when there are $>10$ sources per bin. The confidence limits from \citet{gehrels1986} are adopted otherwise. In both panels, the grey histogram shows the normalized number of sources in a given logarithmic flux density bin and the vertical dashed grey line indicates the $5\sigma$ detection limit of the COSMOS-XS survey. \textbf{Right:} the cumulative fraction of sources with and without radio excess versus flux density, defined to increase towards lower flux densities. The fraction of sources powered by star formation reaches near-unity below $\sim20\,\mu$Jy. Cumulatively, $\sim90\%$ of our sample shows radio emission that is star-formation powered.}
	\label{fig:qexcessagn}
\end{figure*}

Instead of adopting the MLAGN and HLAGN terminology, which includes sources with signs of AGN activity across their full SED, we consider in Figure \ref{fig:qexcessagn} the fractional contribution of sources with and without radio excess. The latter class includes galaxies that exhibit AGN-like activity in their X-ray to mid-infrared SEDs, but show no sign of AGN activity at radio wavelengths. While at fluxes above $\gtrsim100\mu$Jy sources with radio-excess dominate the population, their fractional contribution declines steeply towards lower flux densities, and below $\lesssim20\mu$Jy the contribution of galaxies without any radio-excess is $\gtrsim95\%$. If we adopt the definition that $-$ despite any other AGN signatures $-$ galaxies without any radio-excess are star-forming, this implies that the fraction of star-forming galaxies is nearly unity below $\lesssim20\mu$Jy. Overall, the fraction of sources without AGN signatures in the radio in the COSMOS-XS survey equals $91.2\pm0.8\%$. We verify in Appendix \ref{app:trends} that there is no dependence of any of the AGN diagnostics as function of flux density, indicating that the increased fractional contribution of star-forming sources with decreasing flux density is robust.

\begin{figure*}[!ht]
    \centering
	\includegraphics[width=1.0\textwidth]{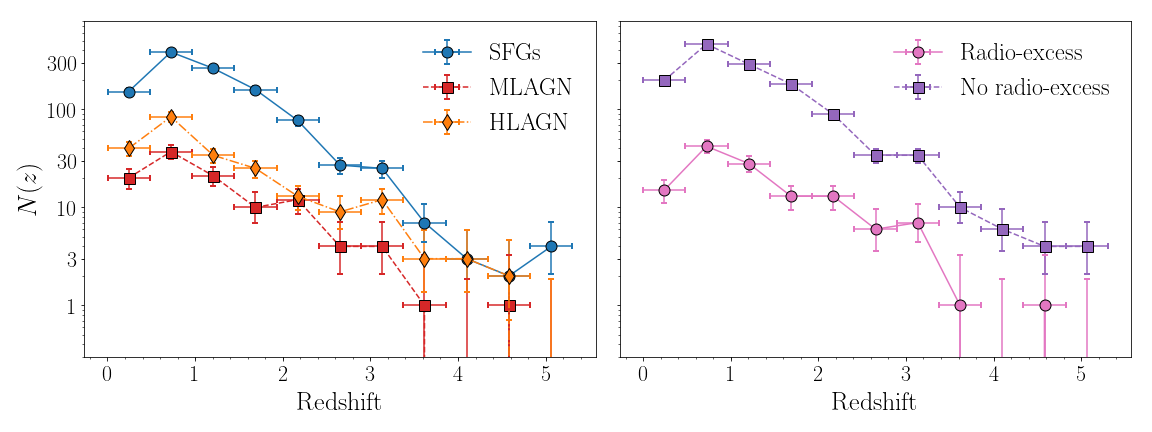}
	\caption{\textbf{Left:} distribution of SFGs, MLAGN and HLAGN with redshift. \textbf{Right:} distribution of radio-excess AGN and sources without radio excess with redshift. Errorbars represent the uncertainties on the counting statistics in each bin. The median redshift of the sample is $z\sim1.0$, and the fractional contribution of the SFGs and various types of AGN remains roughly constant as a function of redshift.}
	\label{fig:redshiftpopulation}
\end{figure*}

In Figure \ref{fig:redshiftpopulation}, we show the distribution of the sample with redshift, and the fractional contribution of each population per redshift bin. The median redshift for all populations is approximately $z\sim1$, illustrating the well-known result that the redshift distribution of radio sources is near-independent of flux density \citep{condon1989}. In addition, the overall fraction of the various source populations remains fairly constant with redshift. This likely indicates that there are no obvious biases in the AGN selection as a function of redshift, which we investigate further for each of the AGN diagnostics individually in Appendix \ref{app:trends}.

\renewcommand{\arraystretch}{1.3}
\begin{deluxetable*}{cccccccccccc}
	\tabletypesize{\small}
	\tablewidth{\textwidth}
	\tablecaption{Euclidean-normalized source counts for the various radio populations.}
	
	\tablehead{
		\multicolumn{3}{c}{\textbf{Flux ($\mu$Jy)}} &
		\multicolumn{4}{c}{\textbf{Completeness}} &
		\multicolumn{1}{c}{\textbf{SFGs}} &
		\multicolumn{1}{c}{\textbf{MLAGN}} &
		\multicolumn{1}{c}{\textbf{HLAGN}} &
		\multicolumn{1}{c}{\textbf{No $q-$excess}} &
		\multicolumn{1}{c}{\textbf{$q-$excess}}
	}
	
	\startdata
	
	$S_\text{low}$ & $S_\text{high}$ & $S_\text{centre}$ & $f_\text{flux}$\tablenotemark{a} & $f_\text{ctrpt}$\tablenotemark{b} & $f_\text{spurious}$\tablenotemark{c} & $C_\text{full}$\tablenotemark{d} & Counts\tablenotemark{e}  & Counts  & Counts  & Counts  & Counts  \\ \tableline \vspace{-2.0ex} \\
    
    2.64 & 4.82 & 3.56 & 0.162 & 0.909 & 0.010 & $7.11_{-1.45}^{+2.37}$ & $0.45_{-0.10}^{+0.16}$ & $0.02_{-0.01}^{+0.01}$ & $0.03_{-0.01}^{+0.02}$ & $0.50_{-0.10}^{+0.17}$ & $0.01_{-0.00}^{+0.01}$ \\
    4.82 & 8.82 & 6.52 & 0.444 & 0.935 & 0.011 & $2.45_{-0.29}^{+0.41}$ & $0.70_{-0.09}^{+0.13}$ & $0.02_{-0.01}^{+0.01}$ & $0.11_{-0.02}^{+0.02}$ & $0.80_{-0.10}^{+0.15}$ & $0.02_{-0.01}^{+0.01}$ \\
    8.82 & 16.12 & 11.92 & 0.828 & 0.938 & 0.017 & $1.28_{-0.10}^{+0.17}$ & $0.94_{-0.09}^{+0.13}$ & $0.08_{-0.02}^{+0.02}$ & $0.17_{-0.03}^{+0.03}$ & $1.11_{-0.10}^{+0.15}$ & $0.07_{-0.02}^{+0.02}$ \\ 
    16.12 & 29.49 & 21.81 & 0.934 & 0.947 & 0.015 & $1.13_{-0.07}^{+0.14}$ & $1.29_{-0.13}^{+0.18}$ & $0.07_{-0.02}^{+0.02}$ & $0.28_{-0.05}^{+0.05}$ & $1.54_{-0.15}^{+0.21}$ & $0.10_{-0.03}^{+0.03}$ \\
    29.49 & 53.94 & 39.88 & 0.940 & 0.900 & 0.020 & $1.17_{-0.13}^{+0.19}$ & $1.16_{-0.19}^{+0.23}$ & $0.36_{-0.08}^{+0.10}$ & $0.59_{-0.12}^{+0.14}$ & $1.70_{-0.26}^{+0.32}$ & $0.41_{-0.09}^{+0.11}$ \\
    53.94 & 98.65 & 72.95 & 1.000 & 0.920 & 0.000 & $1.09_{-0.09}^{+0.20}$ & $0.89_{-0.20}^{+0.24}$ & $0.50_{-0.14}^{+0.16}$ & $0.46_{-0.14}^{+0.16}$ & $1.20_{-0.24}^{+0.29}$ & $0.66_{-0.17}^{+0.20}$ 
    	
	\enddata
	
	\tablenotetext{a}{Flux completeness of the given flux density bin, including the incompleteness resulting from reduced primary beam sensitivity.}
	\tablenotetext{b}{Fraction of radio sources in the given flux density bin assigned a multi-wavelength counterpart with a robustly measured redshift.}
	\tablenotetext{c}{Expected fraction of spurious sources in the given flux density bin.}
	\tablenotetext{d}{Overall completeness correction applied to the bin, as defined in Equation \ref{eq:completeness}, with the propagated uncertainty.}
	\tablenotetext{e}{All the number counts are given in units of $\text{Jy}^{3/2}\ \text{sr}^{-1}$.}
	\label{tab:euclidean}
\end{deluxetable*}
\renewcommand{\arraystretch}{1.0}

\subsection{Euclidean-normalized Number Counts}
In this section, we translate our observed sample, which may be parametrized as $N_i$ radio-detected sources within the $i^\text{th}$ flux density bin $S_{\nu,i}$, into the completeness-corrected Euclidean-normalized number counts. The completeness corrections are required to reconstruct the intrinsic number density of radio sources from our observed sample. Our modus operandi has been to cross-match the 3 GHz detected radio population with various multi-wavelength catalogs, and to use this information to classify this radio sample into AGN and star-forming sources. The main source of incompleteness is the primary beam attenuation, decreasing our sensitivity to faint radio sources towards the edge of the pointing. We additionally correct for spurious detections in the original 3 GHz map, as well as our incompleteness in assigning multi-wavelength counterparts to real radio sources. The magnitude of the former two completeness corrections are detailed in Paper I, and the incompleteness in counterpart association was determined in Section \ref{sec:specz} (see also Figure \ref{fig:counterpartcompleteness}).

In the following, we will assume that the completeness corrections we have derived apply uniformly to the various radio populations - that is, these corrections are a function of observed flux density only, and not of any additional source properties. The full completeness correction $C_i$ applied to the $i^\text{th}$ flux density bin is then given by

\begin{align}
	C_i\left(S_{\nu_i}\right) = \frac{1 - f_\text{spurious}(S_{\nu_i})}{f_\text{flux}(S_{\nu_i}) \times f_\text{ctrpt}(S_{\nu_i})}.
	\label{eq:completeness}
\end{align}

Here $f_\text{spurious}$ is the fraction of spurious sources expected in the given flux density bin, $f_\text{flux}$ is the fractional flux density completeness of our sample, taking into account our declining sensitivity to sources away from the primary beam centre, and $f_\text{ctrpt}$ is the fraction of sources in the given flux density bin for which we have obtained reliable non-radio counterparts. The Euclidean source counts in the $i^\text{th}$ bin are then computed via

\begin{align}
	S_{\nu,i}^{5/2} n(S_{\nu,i}) = \frac{C_i N_i}{\Delta S_{\nu,i}\Omega} S_{\nu,i}^{5/2},
\end{align}

where $\Omega \sim 350 \text{ arcmin}^2$ is the field of view of the S-band survey area, out to 20\% of the maximum primary beam sensitivity and $\Delta S_{\nu,i}$ is the width of the $i^\text{th}$ flux density bin. The normalization with $S_\nu^{5/2}$, applied to the center of the bin, has historically been used, and translates into a flat slope of the number counts with flux density for a fully Euclidean universe. The completeness-corrected Euclidean source counts for the different radio populations are shown in Figure \ref{fig:euclidean} and tabulated in Table \ref{tab:euclidean}. The total uncertainties on our measurements combine uncertainties on the counting statistics with the propagated errors on the various completeness corrections. The effects of cosmic variance are not included in the uncertainties, but we quantify its contribution in Section \ref{sec:cosmicvariance} and overplot the results as shaded regions in the figure. 

We compare our results with the number counts from \citet{smolcic2017b}, who cover a larger area to shallower depths. Focusing first on the sample of MLAGN, HLAGN and clean SFGs (the upper panel of Figure \ref{fig:euclidean}), we observe a good match between the number counts of the two types of AGN between our data and the \citet{smolcic2017b} sample at the flux densities we have in common ($25 \lesssim S_\nu \lesssim 100\, \mu$Jy), whereas we find a slight increase in clean SFGs at the fainter flux densities. This may be explained by cosmic variance (Section \ref{sec:cosmicvariance}), or by uncertainties in the completeness corrections at the faint end of the VLA-COSMOS survey ($\lesssim50\,\mu$Jy). 

In the lower panel of Figure \ref{fig:euclidean}, we compare our number counts for sources with and without radio-excess with the VLA-COSMOS sample. We find an overall agreement, although our fraction of sources without radio-excess at flux densities $\lesssim50\,\mu$Jy is slightly larger than what is found by \citet{smolcic2017b}, similar to our increase in the counts for clean SFGs. The number counts of radio-excess AGN are further in good agreement at the flux densities the two surveys have in common. Combined, this is fully consistent with our results from Paper I, where we find a slight increase in the overall radio number counts compared to the 3 GHz VLA-COSMOS sample. Similar to the above, these differences may be explained by cosmic variance or uncertainties in the completeness corrections. Overall, our 3 GHz source counts are broadly consistent with the VLA-COSMOS data, and as ultimately the population of sources with and without radio excess are used to determine cosmic star-formation rate densities, this agreement is encouraging. 

We additionally compare our results to recent simulations from \citet{bonaldi2018}, who simulate the radio continuum emission from radio-excess AGN and star-forming galaxies. They model the star-forming population by converting galaxy luminosity functions from UV, Lyman-$\alpha$ and IR observations from \citet{cai2013,cai2014} $-$ corrected for dust extinction where necessary $-$ into the star-formation history of the universe (SFHU). As star formation is directly coupled to radio emission, this SFHU is sampled and converted into a radio luminosity function using the local calibration from \citet{murphy2012}, slightly adapted to compensate for the over-prediction of the faint end of the local radio luminosity functions (see the discussion in \citealt{mancuso2015}). For the radio-excess AGN, \citet{bonaldi2018} start with radio luminosity functions for three different types of AGN (steep-spectrum sources, flat-spectrum radio quasars and BL Lacs), which they evolve through cosmic time using evolutionary parameters motivated by the literature. Additionally, different spectral indices are assumed for the three sets of AGN, including a Gaussian distribution around the mean spectral index per class of AGN. Their resulting definitions of the AGN and star-forming populations are largely consistent with our definition of sources with and without radio-excess, respectively, and hence we show the source counts determined from their simulated catalog in the lower panel of Figure \ref{fig:euclidean}. For this, we use the $25\,\text{deg}^2$ simulations at 3 GHz that probe down to flux densities of $10\,\text{nJy}$. Our number counts for star-formation powered sources are in excellent agreement with these simulations across the full range of flux densities we cover, including at the faint end, where we are probing a fully new parameter space. Our number counts for radio-excess AGN are additionally consistent with the predictions from \citet{bonaldi2018}, although the last two bins for radio-excess AGN ($S_\text{3\,GHz} \lesssim 10\,\mu$Jy) lie slightly below the expected value from the simulations. However, our number counts are still consistent with the predictions within the $1\sigma$ scatter due to cosmic variance, which we quantify in the following section.

\subsubsection{Cosmic Variance}\label{sec:cosmicvariance}

To quantify how cosmic variance may influence our observed number counts, we make use of the $5\times5\ \text{deg}^2$ simulations by \citet{bonaldi2018} that model the populations of star-forming and radio-excess galaxies. We draw radio sources from non-overlapping circular regions from their full simulated cosmic volume, whereby we take into account that faint sources can only be recovered in the central regions of the primary beam. For example, a source with a flux density of $10\times\sigma_\text{rms}$ can be detected at $5\sigma$ significance out to the half power point of the primary beam. Such sources are therefore drawn from a circular region with a diameter equal to the FWHM of the S-band primary beam. The brightest sources can be detected in the full field of view of our observations, which equals $\Omega \simeq 350\ \text{arcmin}^2$ out to 20\% primary beam sensitivity. As such, our cosmic variance calculation takes into account that faint sources, while more numerous than brighter ones, can only be detected within a smaller region in our pointing.

\begin{figure*}
	\centering
	\includegraphics[width=.95\textwidth]{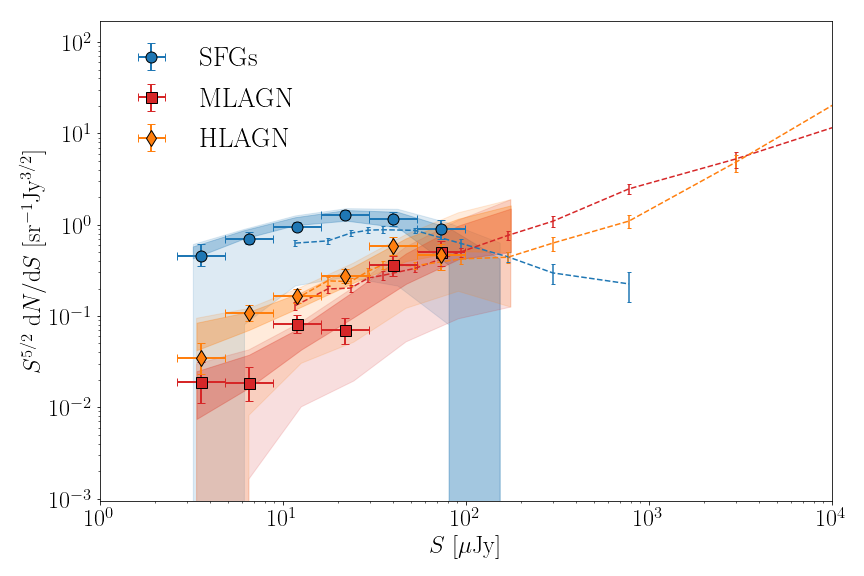} \hfill
	\includegraphics[width=.95\textwidth]{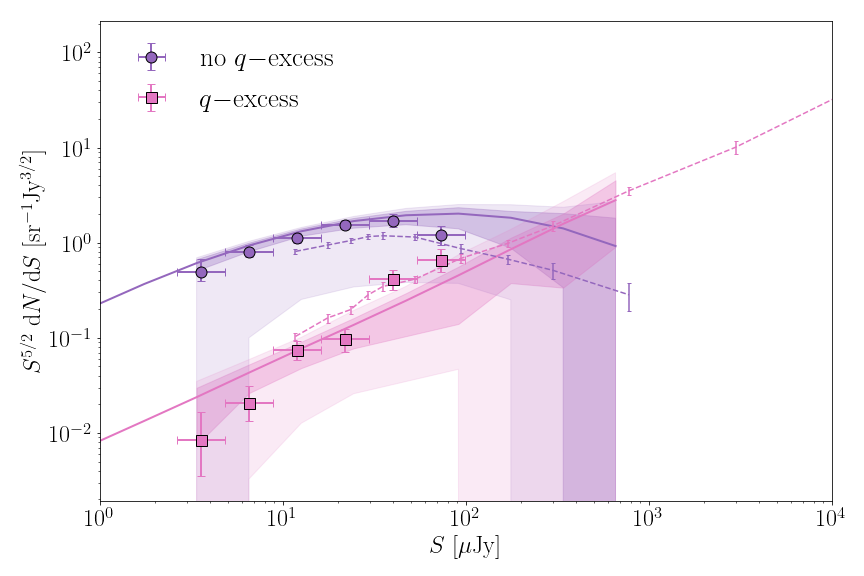}
	\caption{Euclidean-normalized source counts for the faint radio population. \textbf{Upper:} source counts for the moderate-to-high luminosity AGN (HLAGN, orange), low-to-moderate luminosity AGN (MLAGN, red) and clean star-forming galaxies (blue). Our data are represented by the colored markers, and we compare with the recent 3 GHz VLA-COSMOS observations \citep{smolcic2017a,smolcic2017b}, represented by the dashed lines. The expected scatter as a result of cosmic variance is shown by the shaded regions (dark: $1\sigma$, light: $2\sigma$). \textbf{Lower}: source counts for the radio sources with and without radio excess (pink and purple, respectively). The source counts from the \citet{bonaldi2018} simulations are shown by the solid lines. We omit these from the upper plot, as their simulations do not model the MLAGN and HLAGN populations separately.}
	\label{fig:euclidean}
\end{figure*}

For each of the resulting 225 regions, we compute the Euclidean number counts and the corresponding $1\sigma$ and $2\sigma$ confidence intervals, which are shown as the shaded regions in Figure \ref{fig:euclidean}. These confidence intervals hence reflect two effects: at low flux densities the effective field of view of our observations is small, such that the uncertainties resulting from cosmic variance are large. At high flux densities, the effect of cosmic variance is similarly large as bright sources are relatively rare, and this outweighs the increased effective field of view.

As the simulations by \citet{bonaldi2018} do not explicitly model the population of AGN without radio-excess, we combine our observed number counts with their results to obtain an estimate of the effect of cosmic variance for this sub-population of AGN. We extrapolate our measured number counts for clean SFGs, HLAGN and MLAGN to fainter flux densities by fitting a quadratic function in log-space to the measured number counts of SFGs, and a linear function to the MLAGN and HLAGN. We then draw sources from the \citet{bonaldi2018} simulations based on the expected ratios of these three radio populations, and repeat the cosmic variance calculation as for the star-forming and radio-excess samples. We caution that our extrapolations are not based on any physical model, but we judge the cosmic variance results to be robust in the range of flux densities we probe ($\sim2-100\,\mu$Jy) based on the good correspondence between the fits, our data and the shape of the number counts as obtained from the \citet{bonaldi2018} simulations. The typical $1\sigma$ uncertainties as result of cosmic variance are $\sim0.1$ dex for SFGs, and $\sim0.3$ dex for AGN $-$ substantial compared to the formal uncertainties on the derived Euclidean number counts. The additional uncertainty due to cosmic variance must therefore be taken into account when comparing our results to different radio surveys.

\subsection{Optically dark Sources}\label{sec:radioonly}

In total, our radio sample comprises 70 sources that were not matched to a counterpart in any of the three catalogs used for cross-matching. As we expect only $\sim20$ spurious detections, most of the unmatched sources are likely to be real. About half of these radio-only detected sources have Super-deblended counterparts within $2\farcs0$, which is beyond our adopted cross-matching radius of $0\farcs9$. However, visual inspection of these sources at shorter wavelengths shows that the nearby Super-deblended entries are potentially associated to the same source, but exhibit spatial offsets between the infrared ($K_s$ and IRAC $3.6\mu$m) and radio emission, similar to what has been observed at mm-wavelengths (e.g. \citealt{hodge2012}). Nevertheless, a substantial number of detections have $\text{SNR}\gtrsim6$ yet have no counterpart within $\gtrsim5''$, making it likely that these are indeed optically-dark radio sources.

For an additional 53 S-band detected sources we obtained a counterpart in the Super-deblended catalog, but no counterpart at shorter wavelengths. A subset of 23 of these sources was previously catalogued in the $K_s$-band selected catalog by \citet{muzzin2013}, but was excluded from the COSMOS2015 catalog as they are located close to saturated optical stars. As this might affect their optical and near-IR photometry and consequently any derived physical quantities such as their photometric redshifts, we have not further analyzed this sample. The remainder of the sources we solely identify with a counterpart from the Super-deblended catalog are however based on a single detection at 3 GHz in the VLA-COSMOS survey, which we verify in this work through our deeper radio observations. These 30 sources have no existing photometric detections below MIPS $24\,\mu$m, which is a potential sign of obscured AGN at $z\gtrsim6$ (e.g.\ \citealt{rujopakarn2018}) or highly dusty SFGs (e.g.\ \citealt{simpson2014,dudzeviciute2019}), both of which evade detection at short wavelengths. Whereas no robust photometric redshifts for these sources exist due to the lack of optical/near-IR photometry, \citet{jin2018} have employed photometric redshift fitting based on the FIR data available for these sources. While these redshifts are rather uncertain as a result of a degeneracy between redshift and dust temperature (\citealt{jin2018} quote a relative uncertainty of up to $\sim0.1\times(1+z_\text{IR})$ on the derived values), 13 sources are placed beyond $z\geq3$, including 8 sources at $z\geq4$. Additional data are needed to robustly ascertain whether these sources are indeed at such high redshift, but we verified that the sample of $z\geq4$ sources are all roughly consistent with the radio-FIR correlation as obtained in Section \ref{sec:radioexcess}, extrapolated to the FIR-derived redshifts. This indicates that $-$ if these photometric redshifts are correct $-$ the majority of this subsample is likely to have radio emission of a star-forming origin. This is in agreement with our results from Section \ref{sec:countsvsflux}, as we expect $\sim3$ out of 30 sources ($10\%$) to be radio-excess AGN based on their 3 GHz flux densities. \\

We compile a sub-sample of optically dark detections, consisting of sources detected either only at 3 GHz, or cross-matched to a Super-deblended counterpart which in turn was based on a prior position from the VLA COSMOS survey. For the former set, we further ensure that there is no COSMOS2015 counterpart within $2\farcs0$ around any of the detections. This requirement ensures that the remaining sources are indeed `optically dark', as entries in the COSMOS2015 catalog are -- by construction -- detected at short (NIR) wavelengths. The separation of $2\farcs0$ is motivated by the fact that beyond this distance, the distribution of separations between S-band sources and COSMOS2015 counterparts is consistent with being solely the result of chance associations. We further remove all sources with a peak signal-to-noise at 3 GHz below six, to ensure the optically dark sample is robust against spurious detections. We additionally remove any objects flagged due to being located near a bright radio source, leaving us with a sample of 46 sources that we inspect visually in the Subaru/Suprime-cam $i+$, VISTA/Vircam K$_\text{s}$ and \emph{Spitzer}/IRAC $3.6\mu$m bands. From this subset, we then discard radio sources that largely overlap with bright or saturated objects in the optical/near-IR bands, as source blending at these wavelengths is likely to have affected the source detection. The remaining 29 optically dark sources are compiled in Table \ref{tab:opticallydark}.

\def\baselinestretch{1.1}
\begin{deluxetable}{llcccc}
	\tabletypesize{\footnotesize}
	\tablecolumns{5}
	\tablecaption{Optically dark sources detected at 3 GHz.}
	
	\tablehead{
		\colhead{\textbf{RA}} &
		\colhead{\textbf{DEC}} &
		\colhead{\textbf{Flux}\tablenotemark{a}} &
		\colhead{\textbf{Error}\tablenotemark{a}} & 
		\colhead{\textbf{SNR}} & 
		\colhead{\textbf{S17}\tablenotemark{b}}
	}
	
	\startdata
	
	 &  & ($\mu$Jy) & ($\mu$Jy) & & \\ \tableline \vspace{-1.0ex}\\
	
    10$^\text{h}$00$^\text{m}$12.05$^\text{s}$ & 02$^\circ$36$'$48.95$''$ & 33.73 & 0.76 & 44.5 & 1 \\
    10$^\text{h}$00$^\text{m}$00.40$^\text{s}$ & 02$^\circ$36$'$12.40$''$ & 32.66 & 1.72 & 28.3 & 1 \\
    10$^\text{h}$00$^\text{m}$03.82$^\text{s}$ & 02$^\circ$26$'$31.33$''$ & 36.81 & 1.39 & 26.5 & 1 \\
    10$^\text{h}$00$^\text{m}$38.06$^\text{s}$ & 02$^\circ$28$'$06.06$''$ & 20.27 & 0.83 & 24.5 & 1 \\
    09$^\text{h}$59$^\text{m}$58.46$^\text{s}$ & 02$^\circ$30$'$34.58$''$ & 26.06 & 1.07 & 24.3 & 1 \\
    10$^\text{h}$00$^\text{m}$03.28$^\text{s}$ & 02$^\circ$34$'$48.64$''$ & 20.06 & 0.84 & 24.0 & 1 \\
    10$^\text{h}$00$^\text{m}$26.66$^\text{s}$ & 02$^\circ$31$'$26.71$''$ & 16.93 & 1.38 & 21.5 & 1 \\
    10$^\text{h}$00$^\text{m}$54.48$^\text{s}$ & 02$^\circ$34$'$35.86$''$ & 28.08 & 1.90 & 17.7 & 1 \\
    10$^\text{h}$00$^\text{m}$03.27$^\text{s}$ & 02$^\circ$29$'$42.80$''$ & 23.03 & 1.65 & 16.9 & 1 \\
    09$^\text{h}$59$^\text{m}$58.79$^\text{s}$ & 02$^\circ$34$'$57.61$''$ & 15.41 & 1.02 & 15.2 & 1 \\
    10$^\text{h}$00$^\text{m}$19.89$^\text{s}$ & 02$^\circ$33$'$31.45$''$ & 8.56 & 0.57 & 15.1 & 0 \\
    09$^\text{h}$59$^\text{m}$59.81$^\text{s}$ & 02$^\circ$34$'$54.66$''$ & 17.69 & 1.55 & 12.0 & 1 \\
    10$^\text{h}$00$^\text{m}$07.39$^\text{s}$ & 02$^\circ$42$'$03.22$''$ & 59.45 & 3.69 & 12.0 & 0 \\
    10$^\text{h}$00$^\text{m}$20.11$^\text{s}$ & 02$^\circ$39$'$39.99$''$ & 12.01 & 1.02 & 11.8 & 1 \\
    10$^\text{h}$00$^\text{m}$24.44$^\text{s}$ & 02$^\circ$37$'$49.34$''$ & 6.77 & 0.74 & 9.1 & 0 \\
    10$^\text{h}$00$^\text{m}$29.93$^\text{s}$ & 02$^\circ$29$'$18.17$''$ & 5.86 & 0.65 & 9.1 & 0 \\
    10$^\text{h}$00$^\text{m}$09.52$^\text{s}$ & 02$^\circ$26$'$48.42$''$ & 10.07 & 1.13 & 8.9 & 0 \\
    10$^\text{h}$01$^\text{m}$03.04$^\text{s}$ & 02$^\circ$29$'$11.96$''$ & 29.77 & 3.52 & 8.4 & 1 \\
    10$^\text{h}$00$^\text{m}$39.20$^\text{s}$ & 02$^\circ$40$'$52.58$''$ & 27.73 & 2.20 & 7.7 & 0 \\
    10$^\text{h}$00$^\text{m}$25.08$^\text{s}$ & 02$^\circ$26$'$07.27$''$ & 7.62 & 1.00 & 7.6 & 1 \\
    10$^\text{h}$00$^\text{m}$35.34$^\text{s}$ & 02$^\circ$28$'$27.00$''$ & 5.83 & 0.77 & 7.6 & 0 \\
    10$^\text{h}$00$^\text{m}$15.11$^\text{s}$ & 02$^\circ$38$'$17.58$''$ & 6.50 & 0.88 & 7.4 & 0 \\
    10$^\text{h}$00$^\text{m}$04.85$^\text{s}$ & 02$^\circ$35$'$59.51$''$ & 6.21 & 0.85 & 7.3 & 0 \\
    10$^\text{h}$00$^\text{m}$08.64$^\text{s}$ & 02$^\circ$32$'$50.79$''$ & 4.50 & 0.68 & 6.6 & 0 \\
    10$^\text{h}$00$^\text{m}$24.56$^\text{s}$ & 02$^\circ$39$'$11.57$''$ & 6.34 & 0.97 & 6.5 & 0 \\
    10$^\text{h}$00$^\text{m}$41.82$^\text{s}$ & 02$^\circ$25$'$47.13$''$ & 8.80 & 1.35 & 6.5 & 1 \\
    10$^\text{h}$00$^\text{m}$48.46$^\text{s}$ & 02$^\circ$36$'$41.25$''$ & 6.54 & 1.01 & 6.4 & 0 \\
    10$^\text{h}$00$^\text{m}$11.73$^\text{s}$ & 02$^\circ$34$'$25.69$''$ & 4.27 & 0.66 & 6.4 & 0 \\
    10$^\text{h}$00$^\text{m}$34.83$^\text{s}$ & 02$^\circ$28$'$35.83$''$ & 4.41 & 0.73 & 6.1 & 0
	\enddata
    
    \tablenotetext{a}{Peak or integrated flux density and its corresponding uncertainty, based on the analysis in Paper I.}
    \tablenotetext{b}{Boolean flag indicating whether the source is also present in \citet{smolcic2017b}.}
    
	\label{tab:opticallydark}
\end{deluxetable}

We investigate the multi-wavelength properties of these sources by stacking on their radio positions in optical/near-IR bands, as well as at 1.4 GHz. We do not attempt to stack in any of the \emph{Herschel} bands, due to the large PSF at these wavelengths. All bands used for stacking, the corresponding photometry and the limiting magnitudes of the stacks are listed in Table \ref{tab:optistacking}. We perform aperture photometry on the optical/near-IR stacks using a $1\farcs5$ diameter aperture for all bands up to Vista/Vircam $K_s$, and adopt a $3\farcs 0$ diameter aperture for the \emph{Spitzer}/IRAC channels owing to the larger point spread function at these wavelengths. We compute uncertainties on the fluxes by placing the apertures on random locations in the $100\times100$ pixel$^2$ stack, away from the central region and computing the flux within these apertures. The median and standard deviation of these measured aperture fluxes are then taken to be the background and typical flux uncertainty in the stack, respectively.

\begin{figure*}
	\centering
	\includegraphics[width=0.99\textwidth]{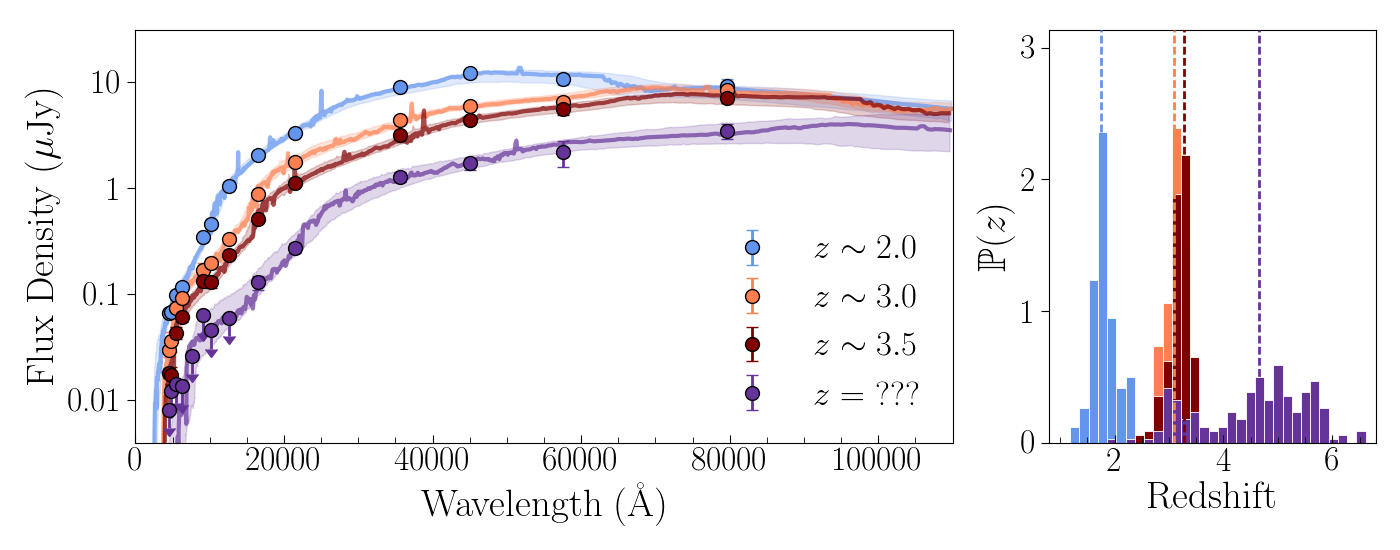}
	\caption{\textbf{Left}: Best fitting SED for a stack of 29 high-redshift radio detections, with a median redshift of $z=1.99$ (blue), $z=2.93$ (orange) and $z=3.42$ (red), as well as the stack for our optically dark sample (purple). The shaded regions indicate the $16^\mathrm{th}-84^\mathrm{th}$ confidence limits around the stacked SED, as obtained from bootstrap re-sampling. \textbf{Right}: The redshift distributions of the bootstrapped optically dark and control samples. The vertical dashed lines indicate the best-fitting redshift for the original stacks. The median redshift of the control samples are fairly well recovered, strengthening the conclusion that the `optically dark' sample consists predominantly of high-redshift ($z\gtrsim 4$) sources.}
	\label{fig:stackedSED}
\end{figure*}

We determine the photometric redshift of the stacked SED through {\sc{EAzY}} \citep{brammer2008}, using the standard set of templates provided with the code. The stack is assigned a best-fitting redshift of $z_\text{stack} = 4.65$, with a $16^\mathrm{th}-84^\mathrm{th}$ confidence interval of $z = (3.22 - 5.63)$. This interval was computed via a bootstrap analysis, whereby we sampled, with replacement, from the 29 optically dark sources, generating a total of 200 bootstrap samples. We then stacked all of these samples, and for each performed aperture photometry and photometric redshift fitting in the same way as for the original stack. The given confidence interval on the best-fitting photometric redshift as such represents the $16^\mathrm{th}-84^\mathrm{th}$ percentile of the bootstrapped redshift distribution. To further verify the robustness of SED-fitting our stacked sample, we additionally stack an identical number of sources that do have optical/NIR photometry, and have a robustly measured photometric or spectroscopic redshift, to see if the average redshift is indeed correctly recovered. For this, we draw 29 sources from our radio sample at random, within three different redshift ranges, requiring only that the sources have no additional COSMOS2015 counterpart within $2\farcs0$ to minimize source blending. Three sets of radio sources were drawn within a redshift of $1.5 \leq z \leq 3.0$ (median redshift $z=1.99$), $2.5 \leq z \leq 4.0$ (median redshift $z=2.93$) and $z \geq 3$ (median redshift $z=3.42$). Our radio sample does not contain sufficiently many sources to further probe higher redshift ranges. The SEDs of both the optically dark and the control samples are shown in Figure \ref{fig:stackedSED}. The recovered photometric redshifts and bootstrapped uncertainties are $z = 1.76_{-0.14}^{+0.28}$, $z=3.10_{-0.21}^{+0.17}$ and $z=3.28_{-0.28}^{+0.10}$ for these three stacked samples, respectively, and hence recover the median redshift of the control sample relatively well. As the redshift probability distribution of the true optically dark sample is substantially broader than that of the control samples, it is likely that it consists of sources across a wider range of redshifts, with a typical value of $z\sim4 - 5$. Nevertheless, if we make the assumption that all sources in the stack lie at the best-fitting photometric redshift, this would imply $\sim90\%$ of our sample at $z\sim5$ is undetected in optical and near-IR photometry. As a sanity check, we compare with the infrared-derived photometric redshifts for the 15 optically dark sources detected only in the Super-deblended catalog from \citet{jin2018}, corresponding to roughly half of the total optically dark sample. This subset has a mean redshift of $z_\text{IR}=3.3$ (with a scatter of $\sigma_z = 1.9$), substantiating the typical high-redshift nature of these detections. We note, however, that these IR-derived redshifts are highly uncertain and therefore mostly indicative, and as such we do not use them in this work. This is further illustrated by the recent work from \citet{jin2019}, who, upon spectroscopic confirmation, find significant differences ($\Delta z \gtrsim 1$) between the true and infrared-derived redshifts for a sample of four dusty galaxies in the COSMOS field. We further note that, at sub-millimeter wavelengths, the typical redshift of the optically dark population is indeed also higher than that of the typical population of sub-millimeter galaxies (e.g. \citealt{simpson2014,dudzeviciute2019}), similar to what we observe here in the radio. 

In addition to stacking in optical/NIR data, the optically dark sources are also detected in the 1.4 GHz stack. The spectral index between the 1.4 and 3 GHz data is found to be $\alpha^{1.4}_{3} = -0.65_{-0.22}^{+0.26}$, consistent with a typical radio spectrum of $\alpha \sim -0.70$. Based on the observed 3 GHz peak flux density of $S_\text{3 GHz} = 10.7 \pm 0.20\ \mu$Jy, we expect $\lesssim 1$ radio-excess AGN to be present in our sample, based on the results in Section \ref{sec:composition}, which in turn implies radio surveys may miss a large fraction of the high-redshift star-forming population when optically dark sources are simply discarded. The consequences of this on the COSMOS-XS-derived cosmic star-formation history will be addressed in a forthcoming paper (Van der Vlugt et al. in prep). If we assume the radio emission from the optically dark sample to be fully powered by star formation, we expect their average star-formation rates to be $\text{SFR} \sim 500 - 1500 \ M_\odot\text{ yr}^{-1}$ when assuming the local value of $q_\text{TIR}$ derived by \citet{bell2003}, and placing the sources at a redshift between $z \sim 3.5 - 6$. Adopting the redshift-dependent conversion from \citet{delhaize2017} instead results in more modest star-formation rates between $150 - 350 \ M_\odot\,\text{yr}^{-1}$, though we stress that both the large spread in the redshift range of the optically dark sample, as well as the unknown value of $q_\text{TIR}$ for typical SFGs above $z\gtrsim3$, are large factors of uncertainty in the derived SFRs. The range of radio luminosities and SFRs spanned by this optically dark sample is shown via the shaded purple region on the COSMOS-XS sensitivity curve in the right panel of Figure \ref{fig:speczfrac_senscurv}.

We lastly investigate the average X-ray properties of the optically dark sample through an X-ray stacking analysis using {\sc{C-STACK}}\footnote{{\sc{C-STACK}} is an online X-ray stacking tool developed by Takamitsu Mijayi, and can be accessed via \url{http://cstack.ucsd.edu/} or \url{http://lambic.astrosen.unam.mx/cstack/}.}, and find that the radio sample is not detected in the X-ray stack, with the count rate in both the soft ($[0.5-2]$ keV) and hard ($[2-10]$ keV) bands being consistent with zero at the $1\sigma$ level. This corresponds to an upper limit of $L_{[0.5-8]\ \text{keV}} \lesssim 2\times10^{43}$ erg/s, assuming a redshift of $z=5.0$ for the stack. Based on this fairly shallow upper limit, we cannot place any constraints on whether the typical source in the stack exhibits X-ray emission suggestive of AGN activity. Nevertheless, the upper limit is consistent with the typical radio-derived SFR for the stack, assuming the FIRC from \citet{delhaize2017} and the X-ray$-$SFR-relations from \citet{symeonidis2014}, further substantiating the notion that the optically dark sample is dominated by high-redshift star-forming galaxies. 

\def\baselinestretch{1.1}
\begin{deluxetable*}{llcccc}
	\tabletypesize{\footnotesize}
	\tablecolumns{6}
	\tablecaption{Data used for stacking of optically dark sample.}
	
	\tablehead{
		\colhead{\textbf{Band}\tablenotemark{a}} &
		\colhead{\textbf{Wavelength ($\mu$m)}} &
		\colhead{\textbf{Depth}\tablenotemark{b}} &
		\colhead{\textbf{Flux Density ($\mu$Jy)}\tablenotemark{c}} &
		\colhead{\textbf{Error ($\mu$Jy)}} &
		\colhead{\textbf{SNR}}
	}
	
	\startdata
	
	Subaru/Suprime-cam $B$ & 0.4458 & 28.6 & 0.003 & 0.003 & 1.2 \\
	Subaru/Suprime-cam $g+$ & 0.4777 & 28.1 & 0.004 & 0.004 & 1.0 \\
	Subaru/Suprime-cam $V$ & 0.5478 & 28.0 & 0.005 & 0.005 & 1.0 \\
	Subaru/Suprime-cam $r+$ & 0.6288 & 28.0 & 0.013 & 0.005 & 2.8 \\
	Subaru/Suprime-cam $i+$ & 0.7683 & 27.3 & 0.022 & 0.009 & 2.5 \\
	Subaru/Suprime-cam $z+$ & 0.9037 & 26.4 & 0.063 & 0.020 & 3.2 \\
	Vista/Vircam $Y$ & 1.0214 & 26.7  & 0.046 & 0.015 & 3.0 \\
	Vista/Vircam $J$ & 1.2534 & 26.5 & 0.060 & 0.019 & 3.2 \\
	Vista/Vircam $H$ & 1.6454 & 26.4 & 0.13 & 0.020 & 6.6 \\
	Vista/Vircam $K_s$ & 2.1540 & 26.0 & 0.27 & 0.030 & 9.1 \\
	\emph{Spitzer}/IRAC CH1 & 3.5634 & 24.2 & 1.29 & 0.15 & 8.9 \\
    \emph{Spitzer}/IRAC CH2 & 4.5110 & 23.8 & 1.72 & 0.22 & 7.7 \\
    \emph{Spitzer}/IRAC CH3 & 5.7593 & 22.7 & 2.20 & 0.60 & 3.7 \\
    \emph{Spitzer}/IRAC CH4 & 7.9595 & 22.8 & 3.47 & 0.57 & 6.1 \\
    VLA/S-band & 3.0 GHz (10 cm) & - & $10.7$ & $0.20$ & 53.5 \\
    VLA/L-band & 1.4 GHz (21 cm) & - & $17.6$ & $3.1$ & 5.7
    
	\enddata
    
    \tablenotetext{a}{References: Subaru/Suprime-cam images from \citet{capak2007}; Vista/Vircam data are UltraVISTA DR3 \citep{mccracken2012}; \emph{Spitzer}/IRAC images are from S-COSMOS \citep{sanders2007}; the VLA L-band mosaic is from \citet{schinnerer2007}.}
    \tablenotetext{b}{For the optical/near-IR bands, this corresponds to the $5\sigma$ limiting magnitude in $1\farcs5$ apertures ($3\farcs0$ for \emph{Spitzer}/IRAC). For the radio data, see the error on the flux density as measure of the depth of the stack.}
    \tablenotetext{c}{For the two radio bands, these are the peak flux densities in $\mu$Jy$\ \text{beam}^{-1}$.}
    
	\label{tab:optistacking}
\end{deluxetable*}

\subsection{Implications for Next-Generation Radio Surveys}
The two largest upcoming radio telescopes are the next-generation VLA (ngVLA) and the Square Kilometer Array (SKA). Both are expected to revolutionize the radio view on galaxy evolution owing to their greatly improved sensitivity and resolution compared to present-day radio interferometers. For example, the ngVLA will be able to reach the COSMOS-XS 3 GHz RMS-sensitivity of $\sim0.53\,\mu$Jy/beam at a similar frequency within an hour of observing time \citep{selina2018}.

Additionally, assuming a spectral index of $\alpha=-0.7$ and an observing frequency of $\sim1$ GHz, the SKA will reach a similar sensitivity as our 3 GHz observations as part of the SKA Wide survey, which is expected to cover an immense area of $\sim10^{3}\,\text{deg}^2$ \citep{prandoni2015}. This is a factor $\sim10^{4}$ larger than our surveyed area, such that a zeroth order scaling of our number counts predicts that the SKA Wide Survey will detect upwards of $\gtrsim 10^{7}$ faint radio sources.

Both our results for the faint radio source counts and predictions from simulations \citep{wilman2008,bonaldi2018} are consistent with a continuous increase in the fraction of sources without radio-excess towards lower flux densities. This is in further agreement with the recent work from \citet{novak2018}, who extrapolate the VLA-COSMOS luminosity functions to predict the relative contributions of SFGs and AGN. Based on a simple power-law extrapolation of our radio number counts below $30\,\mu$Jy, we find that at $\sim1\,\mu$Jy, the fractional contribution of radio-excess sources is $\lesssim1\%$, such that the next-generation radio surveys will be highly sensitive to the faint, star-forming population. In fact, simple flux cuts on the radio-detected sample are likely to be sufficient for obtaining a highly pure sample of star-forming sources.

\section{SUMMARY \& FUTURE}
\label{sec:summary}
We have presented a multi-wavelength analysis of the faint radio population identified in the COSMOS-XS survey, which is the deepest multi-frequency radio survey to date, reaching a $5\sigma$ flux limit of $\sim2.7\,\mu$Jy beam$^{-1}$ within the centre of the 3 GHz image (RMS of $0.53\,\mu\text{Jy}\,\text{beam}^{-1}$). This image, which covers a total area of $\sim350\text{\,arcmin}^2$, is a factor $\sim5$ deeper than the previously deepest radio data over COSMOS at this frequency, and enables the direct detection of the typical star-forming population ($\text{SFR}\lesssim100\,M_\odot\,\text{yr}^{-1}$) out to $z\lesssim3$ (Figures \ref{fig:cosmos} and \ref{fig:speczfrac_senscurv}). To characterize the observed faint radio population, we associated multi-wavelength counterparts to our radio sample by drawing from the several photometric catalogs available over the well-studied COSMOS field. In total, we associated such counterparts to $96.6\%$ of our radio sources. We restrict ourselves to the cross-matches with robustly determined photometric (59\%) or spectroscopic (41\%) redshifts available, accounting for for a total of 1437 sources (93.3\% of the total radio sample). The median redshift of the population is $\langle z \rangle = 0.97 \pm 0.03$, and the sample further includes 51 high-redshift sources with $3.0 \leq z \leq 5.3$.

We separate this faint radio population into star-forming sources and AGN, where the latter are again divided into different subclasses, either based on their radiative luminosities or on an excess in radio emission based on what is predicted from star-formation related emission alone. Sources with moderate-to-high radiative luminosities (HLAGN) exhibit AGN-related emission throughout their multi-wavelength spectral energy distribution, and are identified through strong X-ray emission, AGN-specific mid-IR colors or based on clear AGN-like components from SED-fitting. On the other hand, low-to-moderate luminosity AGN (MLAGN) are identified through an observed lack of star formation, based on red rest-frame UV/optical colors, or an excess in radio emission compared to their full $8-1000\,\mu$m FIR luminosity.

The full catalog of sources detected at 3 GHz, including their multi-wavelength source identifications, and the results from our AGN identification is available with this Paper in standard FITS format. We show a sample of the catalog in Table \ref{tab:catalog}, and elaborate on its contents in Appendix \ref{app:catalog}.

Overall, these multi-wavelength diagnostics identify $23.2\pm1.3\%$ of the COSMOS-XS sample as AGN, with only $8.8\pm0.8\%$ of the radio-detected sources exhibiting AGN-related emission at the observed radio frequencies. Radio emission from the remainder of the sample therefore appears consistent with originating from star formation (Figure \ref{fig:radioexcess}). The incidence of the various types of AGN is a strong function of flux density (Figures \ref{fig:agnwithflux} and \ref{fig:qexcessagn}); the relative contributions of clean SFGs, MLAGN and HLAGN are similar for $S_\text{3\,GHz} \gtrsim 50\,\mu$Jy, but towards lower flux densities the fraction of clean star-forming sources rises rapidly, reaching $\gtrsim80\%$ at $S_\text{3\,GHz} \lesssim 20\,\mu$Jy. Moreover, sources without a radio-excess form the overwhelming majority of our radio sample for $S_\text{3\,GHz} \lesssim 50 \mu$Jy and reach a fraction of $\gtrsim90-95\%$ below $S_\text{3\,GHz} \lesssim 30\,\mu$Jy. Our radio sample is therefore strongly dominated by the star-forming population. Next-generation radio surveys, capable of reaching sensitivities similar to COSMOS-XS in an hour, are therefore ideally suited to probe this faint population and constrain the dust-insensitive star formation history of the universe accordingly.

We additionally observe an interesting subset of radio-detected sources with multi-wavelength counterparts solely at far-infrared wavelengths, or no such counterparts at all. A stacking analysis at optical, near-infrared and radio wavelengths indicates these sources are likely to be high-redshift in nature ($z\sim5$, Figure \ref{fig:stackedSED}). If powered entirely through star-formation, this implies radio-based star-formation rates of hundreds of solar masses per year. As such, these `optically dark' sources may contribute appreciably to cosmic star-formation at high-redshift. 

We further present the completeness-corrected Euclidean-normalized 3 GHz radio number counts within the faint regime ($2\,\mu\text{Jy} \lesssim S_\text{3\,GHz} \lesssim 100 \,\mu\text{Jy}$) for the various radio population in Figure \ref{fig:euclidean}, and show that the COSMOS-XS sample forms a natural extension of the previous literature towards fainter flux densities. Additionally our number counts are in good agreement with recent simulations of the radio sky within the expected scatter of cosmic variance.

Overall, COSMOS-XS provides the deepest multi-frequency radio survey to date, and probes a new, faint parameter space in the radio. In this work $-$ the second paper in the COSMOS-XS series $-$ we have presented the decomposition of the faint radio population into star-forming galaxies and AGN. It will be followed by additional works studying the radio-derived star formation history and AGN accretion history, and additionally, we will leverage the multi-frequency data over COSMOS to further study the radio properties of the COSMOS-XS sample, including the evolution of the far-infrared-radio correlation, and the intrinsic radio spectra of the faint, star-forming population.

\acknowledgements
We wish to thank the anonymous referee for their comments and suggestions which greatly improved this work. The authors wish to thank Chris Carilli for the useful discussions while preparing the proposal and for his help planning the observations, and Mara Salvato for providing us with the COSMOS spectroscopic master catalog. The National  Radio Astronomy Observatory is a facility of the National Science Foundation operated under cooperative  agreement by Associated Universities, Inc. H.A., D.vdV. and J.H. acknowledge support of the VIDI research programme with project number 639.042.611, which is (partly) financed by the Netherlands Organization for Scientific Research (NWO). I.S. acknowledges support from STFC (ST/P000541/1). D.R.\ acknowledges support from the National Science Foundation under grant number AST-1614213. D.R.\ also acknowledges support from the Alexander von Humboldt Foundation through a Humboldt Research Fellowship for Experienced Researchers.



\appendix

\section{AGN Diagnostics from SED Fitting}
\label{app:agnfitter}
We utilize Bayesian Markov Chain Monte Carlo (MCMC) SED-fitting code {\sc{AGNfitter}} \citep{calistrorivera2016,calistrorivera2017} to identify AGN based on a comparison between the integrated luminosities in two sets of components: the direct stellar light and the accretion disk in the wavelength range $0.1-1\,\mu$m, and the MIR-continuum emission from a warm, dusty torus and the stellar-heated dust in the wavelength range $1-30\,\mu$m. As {\sc{AGNfitter}} returns realistic uncertainties on these luminosities through an MCMC-approach, we compare the resulting probability distributions instead of simply comparing the luminosities from the best fit. We write $L_\text{C,p}$ to be the $p^\text{th}$ percentile of the integrated luminosity in component $C$. A source is then identified as an AGN if

\begin{align*}
    \left(L_\text{disk,97.5} \geq L_\text{stellar,2.5}\right) \lor \left(L_\text{torus,97.5} \geq L_\text{cold dust,2.5}\right) \ .
\end{align*}

\begin{figure*}
	\centering
	\includegraphics[width=0.49\textwidth]{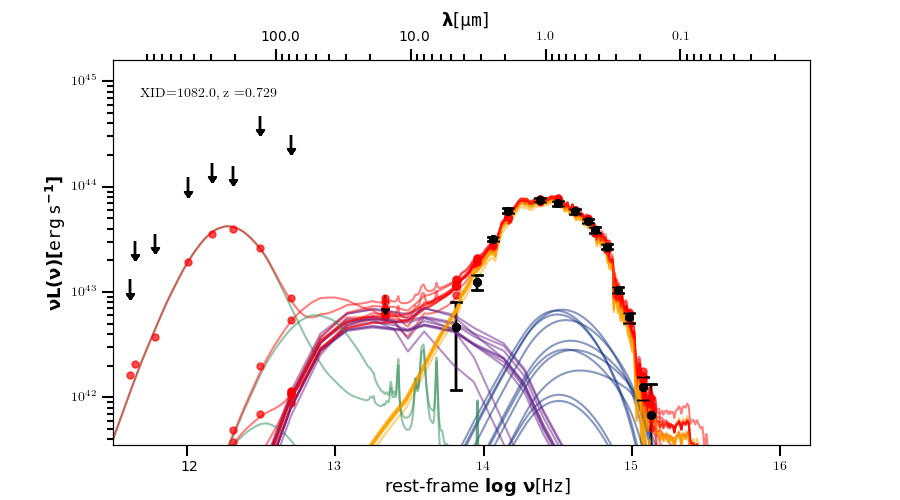}\hfill	\includegraphics[width=0.49\textwidth]{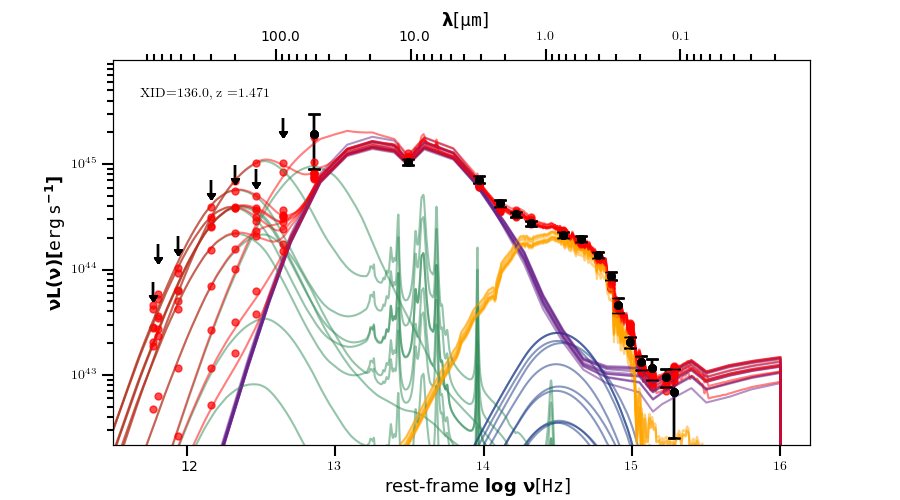}
	\caption{Example SEDs from {\sc{AGNfitter}}. The photometric data are shown through the black points, and upper limits are indicated by downward arrows. The fitted direct stellar contribution is shown in yellow, the accretion disk in blue, the dusty torus in purple and the stellar-heated dust in green. The overall fit is shown in red. A total of 10 random MCMC-realizations are plotted. \textbf{Left:} SED of a source without FIR-photometry, which shows a substantial torus component in the $1-30\mu$m range, and has negligible cold dust component despite having very unconstraining upper limits. Sources such as these are not classified as AGN. \textbf{Right:} SED of a source without reliable ($\geq3\sigma$) FIR-photometry which shows a large torus component and additionally has stringent upper limits on the total FIR-component. Sources such as these are classified as AGN.}
	\label{fig:agnfitterexample}
\end{figure*}

Here $\lor$ denotes the logical or operator, and the $2.5^\text{th}$ and $97.5^\text{th}$ percentiles are equivalent to Gaussian confidence intervals $2\sigma$ below and above the mean, respectively. We slightly modify this procedure for sources without any FIR-photometry, as for this subset of our radio sample the MIR to FIR-SED is nearly fully unconstrained as {\sc{AGNfitter}} does not impose energy balance. As a result, the MIR-continuum of such sources is potentially fully fitted by a torus component, whereas the FIR-luminosity is artificially low, as illustrated in the upper panel of Figure \ref{fig:agnfitterexample}. To avoid identifying such sources as AGN, we define parameter $f_\text{bolo} = L_\text{torus,2.5} / \left(L_\text{torus,2.5} + L_{8-1000\,\mu\text{m}}\right)$, which measures the fractional contribution of the torus to the total FIR-luminosity. Here the luminosity $L_{8-1000\,\mu\text{m}}$ is $-$ by definition $-$ not well-constrained, and therefore we utilize the upper limit on this value as determined empirically in Section \ref{sec:radioexcess}, Figure \ref{fig:radioexcess} (lower panel), where we fit the sensitivity curve of our \emph{Herschel} observations, and hence find a redshift-dependent upper limit on the FIR-luminosity that a source without any FIR-detections can have. Based on visual inspection of the SEDs, as well as the distribution of $f_\text{bolo}$ for sources that do have accurately constrained FIR-luminosities, we then require that sources without FIR-photometry have $f_\text{bolo} \geq 0.20$ in order to be considered an AGN. This, in turn, identifies sources such as the example shown in the lower panel of Figure \ref{fig:agnfitterexample} as an AGN based on a strong torus component and stringent upper limits on the FIR-photometry. Through this procedure, we can then accurately assess whether sources without robust FIR-detections are likely to be AGN based on a substantial torus component, without having to invoke the uncertainties associated with energy balance. Additionally, the separation between SFGs and AGN is done based on physically motivated properties, instead of the comparison of goodness-of-fit.

\section{Comparison to 3 GHz VLA-COSMOS}
\label{app:comparison}
The 3 GHz VLA-COSMOS project \citep{smolcic2017b,smolcic2017a,delvecchio2017} covers the full 2 square degree COSMOS field at an identical frequency, allowing us to compare the multi-wavelength AGN identification for the bright end of our radio sample. Overall, we have 471 radio sources in common (based on cross-matching within $1\farcs0$), subject to two independent analyses. While our overall AGN diagnostics are similar, there are a few differences in the overall analysis. Primarily, we are using improved FIR-photometry taken from the recent Super-deblended catalog \citep{jin2018}, which includes photometry up to 1.2mm, whereas the COSMOS2015 catalog \citep{laigle2016} used in the 3 GHz VLA-COSMOS project contains photometry only up to $500\,\mu$m (\emph{Herschel}/SPIRE). In addition, the deblending procedure employed by \citet{jin2018} is more detailed than previous techniques, and is optimized for highly confused images. Nevertheless, based on the sample both surveys have in common, the overall number of AGN identified is highly similar: we identify 176 out of these 471 sources as AGN, compared to 193 AGN in VLA-COSMOS. The overlap between these two AGN samples is substantial, with 147 of these AGN ($\sim84\%$) being identified in both surveys. The small remaining differences can be explained by the slightly different criteria and multi-wavelength datasets employed in the AGN identification process. We expand upon all of these in the next sections, and summarize the results in Table \ref{tab:agncomparison}.

\subsection{Radio-excess AGN}
\label{app:radioexcess}
In the 3 GHz VLA-COSMOS project, radio-excess AGN are identified through a redshift-dependent threshold in $\log_{10} \left(L_\text{1.4 GHz} / \text{SFR}_\text{IR}\right)$. This quantity is equal to minus $q_\text{TIR}$, up to a constant. The threshold for radio-excess AGN was then determined by considering the scatter towards values below the median in the distribution of $\log_{10} \left(L_\text{1.4 GHz} / \text{SFR}_\text{IR}\right)$, as this regime should predominantly be populated by star-forming galaxies, and will hence trace the intrinsic scatter in the far-infrared radio correlation. Radio AGN should instead only scatter towards larger values. However, since only the observed radio sources are used to determine the spread in the distribution, any upper limits or non-detections will affect the observed scatter. Radio-faint star-forming galaxies in particular are potentially missed, such that the observed spread about the correlation is lower than the intrinsic scatter. As a result, the adopted threshold for classifying AGN is likely to be somewhat conservative, whereby some star-formation powered sources with low $q_\text{TIR}$ will be classified as AGN. In a subsequent paper as part of the 3 GHz VLA COSMOS survey, \citet{delhaize2017} re-calculated the distribution of $q_\text{TIR}$ for star-forming galaxies, now including detection limits in both the far-infrared and radio. We adopted their best-fitting trend for star-forming sources, and classified radio-excess AGN as the sources that are $>2.5\sigma$ offset from the correlation. In addition, we take into account sources without FIR-photometry by comparing the expected infrared luminosity, assuming the radio emission is powered by star formation only, with the \emph{Herschel} sensitivity curves (see the lower panel of Figure \ref{fig:radioexcess}). Nevertheless, the difference in the total number of radio-excess AGN identified in both surveys is small: we find 94 such AGN in COSMOS-XS, whereas there are 102 identified in VLA-COSMOS, among the 471 radio sources in common. This small difference likely reflects the slightly different thresholds adopted.

\subsection{SED-fitted AGN}
\label{app:sedfitting}
In our analysis, we adopted the {\sc{AGNfitter}} SED-fitting code to decompose the full SED of our radio-detected sample into star-forming and AGN components. This code is fundamentally different from {\sc{sed3fit}} used in the 3 GHz VLA-COSMOS project, as it does not adopt energy balance and is based on an MCMC method. As a result, {\sc{AGNfitter}} is designed to distinguish between AGN and star-forming sources not based on a goodness-of-fit test, but instead on physical properties, taking into account the robust uncertainties obtained from MCMC-fitting (see the discussion in Appendix \ref{app:agnfitter}).

Overall, we recover a similar number of AGN through SED-fitting, namely 59 in COSMOS-XS versus 65 in VLA-COSMOS, though only 26 of these are in common between the two radio samples. This difference may most likely be explained by the different fitting codes adopted, in particular whether or not the assumption of energy balance is made, and by the different criteria used for establishing whether a source is AGN. In addition, the slight differences between the two surveys further highlights that there is no clear distinction between AGN and SFGs, and instead indicates that a large fraction of sources is composite in nature. 

\subsection{Red, quiescent AGN}
\label{app:quiescentagn}
In this paper we follow the method of \citet{ilbert2010}, where galaxies are classified as star-forming or quiescent based on their $[\text{NUV}-r^+]$-colors. In particular, both in this work and in \citet{smolcic2017b}, sources are identified as quiescent AGN when $[\text{NUV}-r^+] > 3.5$, where the colors are corrected for dust attenuation. Despite these identical criteria, there is some variation in the AGN identified through this method as in the VLA-COSMOS survey the number of AGN found through red rest-frame UV/optical colors is $\sim58\%$ larger than found in this work (33 in COSMOS-XS versus 52 in VLA-COSMOS, with 28 sources in common). This cannot be explained by our requirement that sources identified through red colors lie below $z\leq2$, as this does not discard any additional red objects from the sample. Nevertheless, in both our and the VLA-COSMOS survey the red, quiescent AGN are also identified through radio-excess in the majority of the cases ($56\%$ in COSMOS-XS versus $64\%$ in VLA-COSMOS), which indicates that sources identified solely through red rest-frame optical/NUV-colors are only a small fraction of the overall AGN population.

\begin{deluxetable}{lccc}
	\tabletypesize{\footnotesize}
	\tablecolumns{4}
	\tablewidth{0.45\textwidth}
	\tablecaption{Comparison of the COSMOS-XS AGN sample with the 3 GHz VLA-COSMOS survey, based on 471 sources in common}
	
	\tablehead{
		\colhead{\textbf{Diagnostic}} &
		\colhead{\textbf{COSMOS-XS}} &
		\colhead{\textbf{VLA-COSMOS}} & 
		\colhead{\textbf{Overlap}}
	}
	
	\startdata
	
	X-ray & 61 & 61 & 56  \\
	IRAC & 20 & 26 & 20 \\
	SED-fitting & 59 & 65 & 26 \\
	Radio-excess & 116 & 102 & 71 \\
	$[\text{NUV}-r^+]$ & 33 & 52 & 28 \\
	\tableline
	\textbf{Overall\tablenotemark{a}} & 190 & 193 & 154
	\enddata
    \tablenotetext{a}{Represents the total number of AGN identified across all five diagnostics in the two surveys.}
	\label{tab:agncomparison}
\end{deluxetable}

\section{Incompleteness in the Multi-wavelength Photometry}
As we utilize multi-wavelength photometry spanning X-ray to radio wavelengths in order to identify AGN, this identification may be affected by incompleteness in the various datasets we employ. We here study the effect this has on our sample of AGN.

\subsection{X-ray AGN}
\label{app:hlagn}
We identify moderate-to-high luminosity AGN (HLAGN) through a combination of X-ray, mid-IR and SED-fitting techniques. Our X-ray data in particular are relatively shallow $-$ we are already incomplete to X-ray sources with a luminosity of $L_\text{[0.5 - 8] keV} \sim 10^{42} \text{ erg s}^{-1}$ beyond $z\sim0.5$ (see also Figure 7 in \citealt{marchesi2016}). While we hence cannot place any direct constraints on the X-ray luminosities of sources beyond this redshift, we employ X-ray stacking to investigate the typical X-ray properties of our clean, star-forming galaxy sample, making use of X-ray stacking tool {\sc{C-STACK}}. By stacking in four redshift bins between $0.5 \lesssim z \lesssim 2.5$, we find that the average X-ray luminosities of this clean SFG sample are fully consistent with the star formation $-$ X-ray luminosity relations from \citet{symeonidis2014}. We will further address this, as well as the stacked X-ray properties for the different classes of AGN, in a forthcoming paper. Overall, this implies that our sample of clean SFGs is likely minimally contaminated by X-ray AGN.

\subsection{MIR AGN}
\begin{figure}
	\centering
	\hspace*{-0.65cm}
	\includegraphics[width=0.55\textwidth]{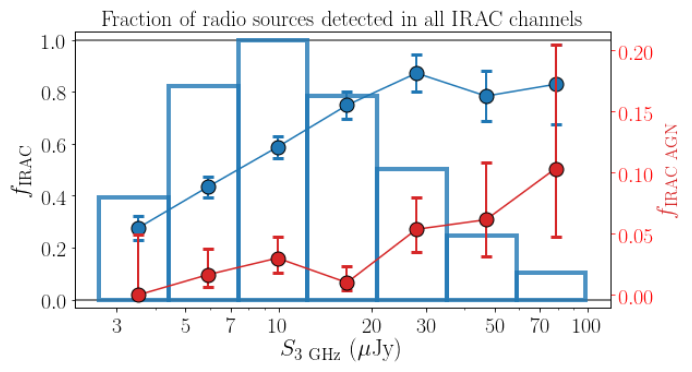}
	\caption{The fraction of sources with measured IRAC flux densities in all four channels as a function of 3 GHz flux density (blue line and points). The background histogram shows the overall distribution of radio flux densities, and the red line - corresponding to the right ordinate axis - denotes the fraction of MIR-AGN we expect when correcting for the incompleteness in IRAC photometry. Similar to our overall results for HLAGN, the fraction of MIR-AGN decreases towards fainter flux densities, even after correcting for the incompleteness.}
	\label{fig:IRACincompleteness}
\end{figure}

In our analysis, we identify 28 sources as mid-IR AGN based on their \emph{Spitzer}/IRAC colors, following the criteria from \citet{donley2012} for sources with $z \leq 2.7$. However, we only have IRAC photometry in all four channels for $\sim60\%$ of our radio sample, which will cause us to miss some MIR-AGN simply based on lack of photometry. In Figure \ref{fig:IRACincompleteness} we show the fraction of radio sources with with IRAC photometry as a function of 3 GHz flux density, which decreases to $\sim30\%$ for our lowest flux density bin ($\sim4\,\mu$Jy).\footnote{We limit this analysis to the 1390 radio sources (96.7\%) with $S_\text{3 GHz} \leq 100\,\mu$Jy due to poor statistics at the brighter end. This includes 25 of the in total 28 MIR-AGN.} The total number of MIR-AGN expected in a given radio flux density bin is then the observed number divided by the fraction of sources with full IRAC photometry. The corrected fraction of MIR AGN is shown with the red datapoints in Figure \ref{fig:IRACincompleteness}. After applying this completeness correction, the fraction of MIR-AGN is still decreasing towards lower radio flux densities, which is a trend we observe for HLAGN in general. Overall, we expect to find $53 \pm 14$ MIR-AGN in our sample of radio sources, after correcting for the IRAC incompleteness, which implies we are missing $28 \pm 8$ additional MIR-AGN. However, as the majority ($\sim80\%$) of IRAC AGN are also identified through either our X-ray or SED-fitting criteria, we expect to miss only $6 \pm 2$ AGN based on our incomplete IRAC photometry, accounting for only $\sim3\%$ of the total number of HLAGN. We are therefore not limited by incompleteness in the MIR-photometry in our identification of (HL)AGN.

\subsection{Radio-excess AGN}
\label{app:firphot}
 Radio-excess AGN are identified through an excess in radio emission compared to what is expected based on their total FIR-luminosity, with this additional radio emission being ascribed to AGN activity. While our full sample is $-$ by definition $-$ detected at 3 GHz, we only have FIR-photometry for $\sim50\%$ of our radio sources, which complicates the determination of accurate total ($8-1000\,\mu$m) infrared luminosities. In order to identify radio-excess AGN, we therefore assume energy balance between the UV to near-IR data and the longer wavelength emission, as the photometry at short wavelengths is typically better constrained. We test the reliability of the FIR-luminosities derived in this way by artificially removing the FIR-photometry of sources that do have detections at such wavelengths and subsequently re-fitting their SEDs with {\sc{magphys}}. For this, we limit ourselves to the 1371 radio sources that we cross-matched with both COSMOS2015 and the Super-deblended catalog. Among these sources, we identified 118 radio-excess AGN (103 and 61 via the normal and `inverse' radio excess criteria, respectively, with substantial overlap; see Figure \ref{fig:qexcessagn}).

After removing all FIR-detections, we recover a total of 143 radio-excess AGN instead (118 and 71 sources through the regular and inverse criteria, respectively -- note that we use the empirical detection limit in Figure \ref{fig:radioexcess} for the latter, as before), $\sim20\%$ larger than the number of AGN we find when we do include available FIR-photometry. This indicates that the energy balance {\sc{magphys}} employs works well for the majority of sources, as even without far-infrared information accurate FIR-luminosities, and hence $q_\text{TIR}$, are predicted. We further note that the `inverse' radio-excess criterion we apply requires a priori information on the far-infrared properties of a subset of our radio sample, as these are used to determine an appropriate sensitivity limit for the various \emph{Herschel} photometric bands. When removing all FIR-information, we can no longer directly apply this criterion. Using only the standard radio-excess diagnostic, we recover a total of 118 AGN -- equal to the number identified through the combined radio-excess and `inverse' radio-excess criteria when far-infrared data are included. Overall, we therefore conclude that the lack of far-infrared photometry for half of our radio sample does not substantially impede our classification of radio-excess AGN.

\subsection{Summary}
\label{app:trends}
\begin{figure*}
	\centering
	\hspace*{-0.65cm}
	\includegraphics[width=\textwidth]{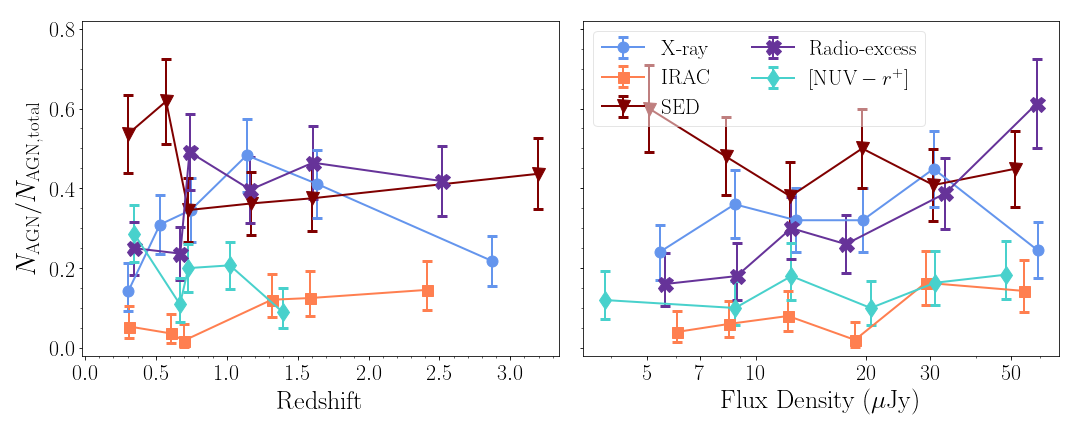}
	\caption{\textbf{Left:} the fractional contribution of the different AGN diagnostics (X-ray emission, \emph{Spitzer}/IRAC colors, SED-fitting, radio-excess and red rest-frame NUV/optical colors) to the overall sample of AGN as a function of redshift. The data were binned into six fixed redshift ranges, containing an approximately equal number of AGN in total. The redshift value of each datapoint marks the median value per bin, and the errorbars represent Poissonian uncertainties. Note that the fractional contribution sums to a number greater than unity, as a single AGN may be identified through multiple diagnostics. The lack of strong trends with redshift indicate that we do not suffer from substantial biases as function of redshift (e.g. luminosity incompleteness) in our identification of AGN. \textbf{Right:} the fractional contribution of the different AGN diagnostics to the overall sample of AGN as a function of flux density. The data follow a similar binning as used in the left panel. With exception of the radio-excess AGN, which we established show a real and significant trend with flux density in the main text, no trends of the other AGN diagnostics are seen to correlate with radio flux density. This indicates that our various methods of AGN identification are reliable across the full range of fluxes probed in the COSMOS-XS survey.} 
	\label{fig:AGNtrends}
\end{figure*}

In the previous subsections, we have established that incompleteness issues are not expected to have a substantial effect on the AGN classification in this work. What remains is then to show that any of the trends we see for the SFGs and various types of AGN with either flux density (Figures \ref{fig:agnwithflux}, \ref{fig:qexcessagn} and \ref{fig:euclidean}) or redshift (Figure \ref{fig:redshiftpopulation}) are not caused by any of the AGN diagnostics being more efficient in identifying sources within a given flux density or redshift interval. In Figure \ref{fig:AGNtrends}, we therefore plot the fractional contribution of each of these diagnostics to the total AGN counts versus redshift (left panel) and flux density (right panel). No strong trends are visible in either of the panels, which indicates that we are not biased towards finding a specific subset of AGN in a given bin of our redshift- and flux density space.\footnote{We, however, caution that one cannot robustly disentangle whether a trend with redshift is a physical effect or the result of a selection bias. Nevertheless, a lack of any given trend substantiates that no such bias is present.} In particular, the fractional contribution of the various types of AGN is a near-constant function of flux density, which substantiates our results in Section \ref{sec:composition} that AGN make up a comparatively smaller fraction of the faint radio population.

\section{FINAL COUNTERPART CATALOG}
\label{app:catalog}

The catalog of sources detected at 3 GHz, including their multi-wavelength source identifications, is available with this Paper in standard FITS format. We show a sample of the catalog in Table \ref{tab:catalog}, and elaborate here on its contents.

\begin{enumerate}
	\setlength{\itemsep}{-0.05cm}
	\item \emph{Column 1.} Source ID, equal to the ID assigned in Paper I.
	\item \emph{Columns 2-3.} Radio coordinates of the source as determined by \texttt{PyBDSF}. 
	\item \emph{Columns 4-5}. Optimal redshift of the radio source, including a boolean indicating whether it is spectroscopic or photometric.
	\item \emph{Columns 6-11}. Flux densities and errors of the source at, respectively, 1.4, 3 and 10 GHz, in $\mu$Jy. In the absence of a radio counterpart at 1.4 or 10 GHz, both the flux and error are set to -99.
	\item \emph{Columns 12-14}. ID of the source in, respectively, the Super-Deblended catalog, COSMOS2015 and the $i-$band selected catalog. If no counterpart was found, this value is set to -99.
	\item \emph{Column 15}. Boolean indicating whether the source is flagged as `potentially spurious' based on the discussion in Section \ref{sec:radiodata}, in which case it is set to True.
	\item \emph{Columns 16-18}. The rest-frame 1.4 GHz luminosity of the source and its $16^\text{th}$ and $84^\text{th}$ percentiles (equivalent to $1\sigma$ confidence intervals), using the measured spectral index if available, or a fixed value of $\alpha=-0.7$ otherwise. The luminosity is given in $\text{W}\,\text{Hz}^{-1}$.
	\item \emph{Columns 19-21}. The value of the radio-FIR correlation parameter $q_\text{TIR}$ as defined in equation \ref{eq:qtir}, and its $16^\text{th}$ and $84^\text{th}$ percentiles.
	\item \emph{Column 22}. Boolean indicating whether the source is identified as an X-ray AGN, in which case it is set to True.
	\item \emph{Column 23}. Boolean indicating whether the source is identified as an AGN through its mid-IR colors, by means of the \citet{donley2012} wedge.
	\item \emph{Column 24}. Boolean indicating whether the source is identified as an AGN based on SED fitting.
	\item \emph{Column 25}. Boolean indicating whether the source is identified as an SED-AGN based on a MIR torus component.
	\item \emph{Column 26}. Boolean indicating whether the source is identified as an SED-AGN based on a UV/optical accretion disk component.
	\item \emph{Column 27}. Boolean indicating whether the source is identified as an AGN based on red rest-frame near-UV and optical colors.
	\item \emph{Column 28}. Boolean indicating whether the source is identified as an AGN based on an excess in radio emission from what is expected from the radio-FIR correlation, with FIR luminosities calculated using {\sc{magphys}}.
	\item \emph{Column 29}. Boolean indicating whether the source is identified as an AGN based on an excess in radio emission from what is expected from the radio-FIR correlation, in the absence of \emph{Herschel} FIR-photometry, through comparison with the detection limit of \emph{Herschel}.
	\item \emph{Column 30}. Boolean indicating whether the source is identified as an HLAGN, which is True when either of columns $22-26$ are True.
	\item \emph{Column 31}. Boolean indicating whether the source is identified as an MLAGN, which is True when all of columns $22-26$ are False and either of columns $27-29$ are True.
	\item \emph{Column 32}. Boolean indicating whether the source is a radio-excess AGN, which is True when either of columns $28-29$ is True.
	\item \emph{Column 33}. Boolean indicating whether the source is an AGN without radio excess, which is True when either of columns $22-27$ are True and both columns $28-29$ are False.
\end{enumerate}

\renewcommand{\arraystretch}{1.3}
\begin{deluxetable*}{cccccccccc}

    \tabletypesize{\footnotesize}
	\tablecolumns{10}
	\tablewidth{\textwidth}
	\tablecaption{Sample of the COSMOS-XS Multi-wavelength and AGN catalog}
	
	\tablehead{
		\colhead{\textbf{ID}} &
		\colhead{\textbf{RA}} &
		\colhead{\textbf{DEC}} &
		\colhead{\textbf{Redshift}} &
		\colhead{\textbf{Spec$-z$}} &
		\colhead{\textbf{ }} &
		\colhead{\textbf{$\mathbf{S_\text{3\,GHz}}$}} &
		\colhead{\textbf{$\mathbf{\sigma_\text{3,GHz}}$}} &
		\colhead{\textbf{ }} &
		\colhead{\textbf{No-$q$E AGN}}
	}
	
	\startdata
    
    & [deg] & [deg] & &  & & [$\mu$Jy] & [$\mu$Jy] & &  \\ \tableline \vspace{-1.0ex} \\
    
    CXS J100016.55+023309.17 & 150.06894 & 2.55255 & 4.36 & 0 & ... & 4.95 & 0.59 & ... & 1 \\
    CXS J095952.41+023148.56 & 149.96837 & 2.53016 & 1.60 & 0 & ... & 8.26 & 1.39 & ... & 0 \\
    CXS J100034.78+022849.87 & 150.14493 & 2.48052 & 1.56 & 0 & ... & 10.56 & 0.71 & ... & 0 \\
    CXS J100036.03+023937.91 & 150.15012 & 2.66053 & 1.22 & 0 & ... & 507.22 & 1.16 & ... & 0 \\
    CXS J100019.40+022936.63 & 150.08083 & 2.49351 & 0.21 & 0 & ... & 5.22 & 0.62 & ... & 0 \\
    CXS J100057.22+023644.73 & 150.23842 & 2.61243 & 1.77 & 0 & ... & 10.11 & 1.56 & ... & 0 \\
    CXS J100028.30+024008.46 & 150.11790 & 2.66902 & 0.35 & 1 & ... & 13.27 & 1.38 & ... & 0 \\
    CXS J100000.45+022739.28 & 150.00188 & 2.46091 & 0.73 & 0 & ... & 12.79 & 1.37 & ... & 1 \\
    CXS J100016.38+023620.36 & 150.06823 & 2.60565 & 2.26 & 0 & ... & 3.90 & 0.69 & ... & 0 \\
    CXS J100015.53+023742.11 & 150.06469 & 2.62836 & 0.69 & 0 & ... & 11.87 & 1.31 & ... & 0

    \enddata

\tablecomments{The contents of the COSMOS-XS multi-wavelength catalog, including a description of the various columns, are elaborated in the text.}

\label{tab:catalog}
\end{deluxetable*}
\renewcommand{\arraystretch}{1.0}

\bibliographystyle{apj}
\bibliography{references}

\end{document}